\documentclass[final,times,twocolumn]{elsarticle}
\usepackage{graphicx,amssymb,amsmath}
\usepackage{multirow,dcolumn,bm,latexsym,soul}
\newcolumntype{K}[1]{>{\centering\arraybackslash}p{#1}}

\journal{Physics of the Dark Universe}
\begin{document}
\newcommand{\be}{\begin{equation}}
\newcommand{\ee}{\end{equation}}
\newcommand{\bq}{\begin{eqnarray}}
\newcommand{\eq}{\end{eqnarray}}
\begin{frontmatter}

\title{Constraining alternatives to a cosmological constant: generalized couplings and scale invariance}
\author[inst1,inst2]{C. B. D. Fernandes}\ead{up201706002@fc.up.pt}
\author[inst1,inst3]{C. J. A. P. Martins\corref{cor1}}\ead{Carlos.Martins@astro.up.pt}
\author[inst1,inst2]{B. A. R. Rocha}\ead{up201604851@fc.up.pt}
\address[inst1]{Centro de Astrof\'{\i}sica da Universidade do Porto, Rua das Estrelas, 4150-762 Porto, Portugal}
\address[inst2]{Faculdade de Ci\^encias, Universidade do Porto, Rua do Campo Alegre 687, 4169-007 Porto, Portugal}
\address[inst3]{Instituto de Astrof\'{\i}sica e Ci\^encias do Espa\c co, CAUP, Rua das Estrelas, 4150-762 Porto, Portugal}
\cortext[cor1]{Corresponding author}

\begin{abstract}
We present a comparative analysis of observational low-redshift background constraints on three candidate models for explaining the low-redshift acceleration of the universe. The generalized coupling model by Feng and Carloni and the scale invariant model by Maeder (both of which can be interpreted as bimetric theories) are compared to the traditional parametrization of Chevallier, Polarski and Linder. In principle the generalized coupling model, which in vacuum is equivalent to General Relativity, contains two types of vacuum energy: the usual cosmological constant plus a second contribution due to the matter fields. We show that the former is necessary for the model to agree with low-redshift observations, while there is no statistically significant evidence for the presence of the second. On the other hand the scale invariant model effectively has a time-dependent cosmological constant. In this case we show that a matter density $\Omega_m\sim0.3$ is a relatively poor fit to the data, and the best-fit model would require a fluid with a much smaller density and a significantly positive equation of state parameter. 
\end{abstract}
\end{frontmatter}


\section{Introduction}
\label{sect1}

The search for the physical mechanism underlying the observed low-redshift acceleration of the universe is the most compelling goal of modern fundamental cosmology. A number of theoretical possibilities can be envisaged in principle, whose observational consequences are being explored \cite{Copeland,Frieman,Joyce}.

The simplest possibility is a cosmological constant: this has the minimal number of additional parameters and indeed is, broadly speaking, in agreement with the currently available data (despite several recent observational hints of inconsistencies). Nevertheless, the observationally inferred value is theoretically unexpected, and reconciling the two would require fine-tuning or some other radical departure from current knowledge. The next-to-simplest possibility would be one (or more) additional dynamical degrees of freedom---particularly scalar fields, which are known to be among Nature's building blocks. Indeed, many (perhaps most) phenomenological dark energy studies explicitly or implicitly assume that the source of the dark energy is a canonical scalar field. Finally, more radical (or, arguably, epicyclic) approaches rely on modifications of the behaviour of gravity. Each of these alternative paradigms will have its observational fingerprints, which one can look for in the ever-improving available data \cite{Huterer}.

Our goal in this work is to present a comparative study of the observational constraints on three classes of models. Two of these are recently proposed models: the generalized coupling model by Feng and Carloni \cite{Feng} and the scale invariant model by Maeder \cite{Maeder}. Both of these models can be interpreted as bimetric theories. As a benchmark for the more standard models we use the traditional phenomenological parametrization of Chevallier, Polarski and Linder (henceforth CPL) \cite{Chevallier,Linder}. All three models have common parameters (specifically, the matter density parameter, $\Omega_m$) but also some specific ones, and a comparative analysis using a common data set is therefore interesting.

In this work we take all three models at face value and phenomenologically constrain them using low-redshift background cosmology data, further described in the next section. The plan of the rest of the paper is as follows. We start in Sect. \ref{sect0} with a brief summary of the data and statistical analysis methodology we use. After this, in the following three sections we introduce each of the three models and present the constraints obtained from the aforementioned data sets, under various assumptions. Specifically, the CPL model is discussed in Sect. \ref{sect2}, the generalized coupling model in Sect. \ref{sect3}, and the scale invariant model in Sect. \ref{sect4}. Finally in Sect. \ref{sect5} we discuss our results and present some conclusions.


\section{Data and methods}
\label{sect0}

We start with a short description of our analysis methodology and of the datasets that we will be used in the analysis. We follow a standard likelihood analysis (see for example \cite{Verde}), with the likelihood defined as
\be
{\cal L}(p)\propto\exp{\left(-\frac{1}{2}\chi^2(p)\right)}.
\ee
As has already been mentioned, we use low-redshift background cosmology data, specifically from supernovas and Hubble parameter data. The two datasets are independent, so the total chi-square is the sum of the two, $\chi^2=\chi^2_{SN}+\chi^2_{HZ}$. Our main observable in both cases with be the re-scaled Hubble parameter, which we define here for convenience
\be
E(z)=\frac{H(z)}{H_0}\,;
\ee
in the following sections we will in general write the Friedmann equation for each model in terms of $E(z)$.

Specifically, for the supernovas we use the recent Pantheon catalogue of Type Ia supernovas available in \cite{Riess}, including its covariance matrix. In passing, we note that the reliability of this data has recently been questioned \cite{Steinhardt}. In this case the chi-square can therefore be written, in the general form
\be
\chi^2_{SN}(p)=\sum_{i,j}\left(E_{obs,i}-E_{model,i}(p)\right)C_{ij}^{-1}\left(E_{obs,j}-E_{model,j}(p)\right)\,,
\ee
where $p$ symbolically denotes the parameters for each of the models and $C$ is the covariance matrix of the dataset. Note that this analysis is independent of the Hubble constant, $H_0$.

For the Hubble parameter we use the compilation listed in \cite{Farooq}, which includes 38 measurements up to redshift $z\sim2.36$. In this case the measurements can be assumed to be independent (i.e., the covariance matrix is trivial), but in order to do the analysis in terms of $E(z)$ and thus combine the two datasets in the likelihood one needs to marginalize over the value of the Hubble constant. In fact, this can be done analytically, following the procedure detailed in \cite{Basilakos}. Specifically, we compute three separate quantities
\be
A(p)=\sum_{i}\frac{E_{model,i}^2(p)}{\sigma^2_i}
\ee
\be
B(p)=\sum_{i}\frac{E_{model,i}(p) H_{obs,i}}{\sigma^2_i}
\ee
\be
C(p)=\sum_{i}\frac{H_{obs,i}^2}{\sigma^2_i}
\ee
where the $\sigma_i$ are the uncertainties in observed values of the Hubble parameter. The the chi-square has the following form
\be
\chi^2(p)=C(p)-\frac{B^2(p)}{A(p)}+\ln{A(p)}-2\ln{\left[1+Erf{\left(\frac{B(p)}{\sqrt{2A(p)}}\right)}\right]}
\ee
where $Erf$ is the Gauss error function and $\ln$ is the natural logarithm.

The analysis is done on a grid, for the parameters described in each case in the following sections; since we are only dealing with background cosmology, there is no computational need for a full MCMC analysis. Confidence levels are then identified (in terms of the corresponding $\Delta\chi^2$) with standard numerical tools. We have explicitly verified that the grid sizes that have been used are sufficiently large for the results presented in the following sections not to be affected by these sizes; moreover, the following section will also present an explicit validation test of our code for the supernova data.

Finally, we note that since we will only be concerned with low-redshift data $z<2.5$ we ignore the radiation density in subsequent chapters. This simplifying assumption has no significant impact in our results. Also, we will work in units where the speed of light is set to $c=1$.


\section{Standard cosmology: the CPL model}
\label{sect2}

In the CPL parametrization the dark energy equation of state parameter is assumed to have the form \cite{Chevallier,Linder}
\be
w(z)=\frac{p(z)}{\rho(z)}= w_0+w_a\frac{z}{1+z}\,,
\ee
where $w_0$ is its present value while $w_a$ quantifies its possible evolution in time (or, explicitly, redshift). This is a phenomenological approach, in the sense that it is not intended to mimic a particular dark energy model, but aims to describe generic departures from the $\Lambda$CDM behaviour, which naturally corresponds to $w_0=-1$ and $w_a=0$. In principle it allows for both canonical and phantom fields, since there is no restriction on the two model parameters, at least on purely mathematical grounds.

\begin{figure*}
\begin{center}
\includegraphics[width=3.2in,keepaspectratio]{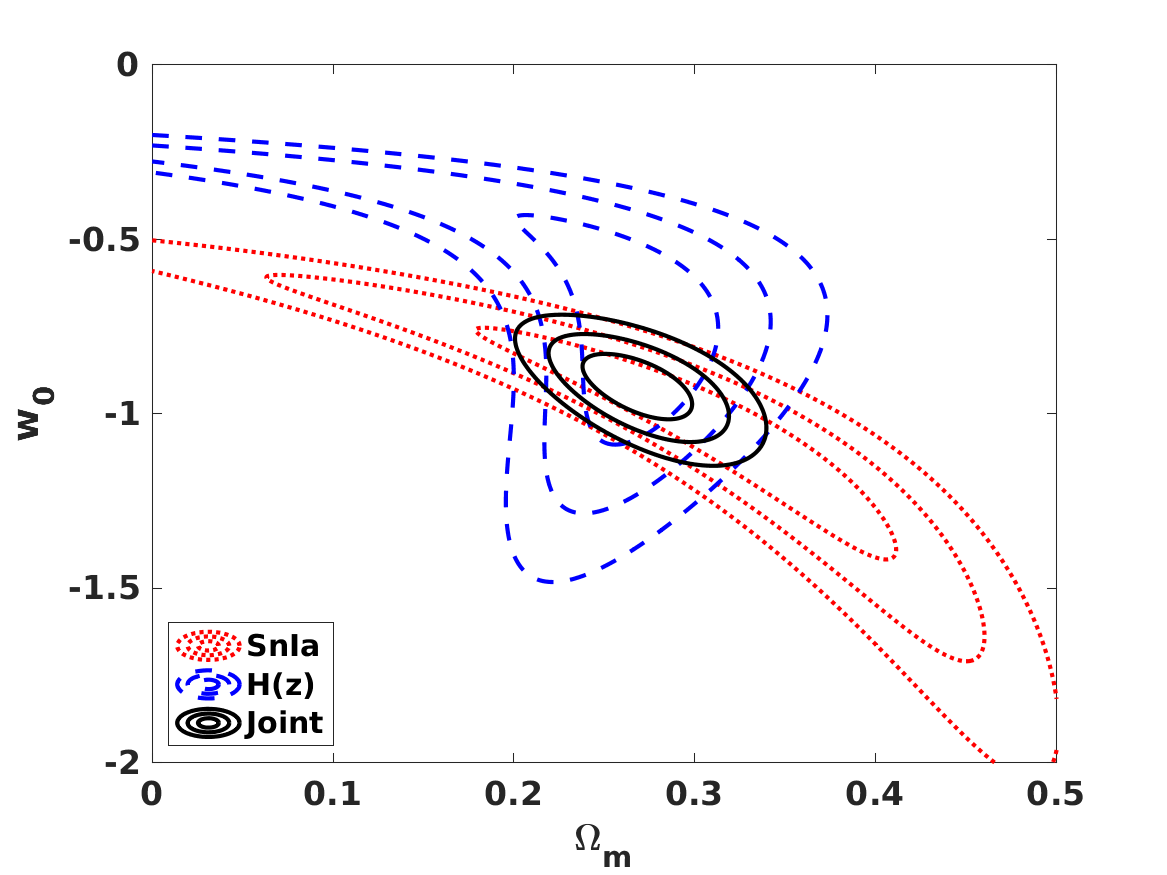}\\
\includegraphics[width=3.2in,keepaspectratio]{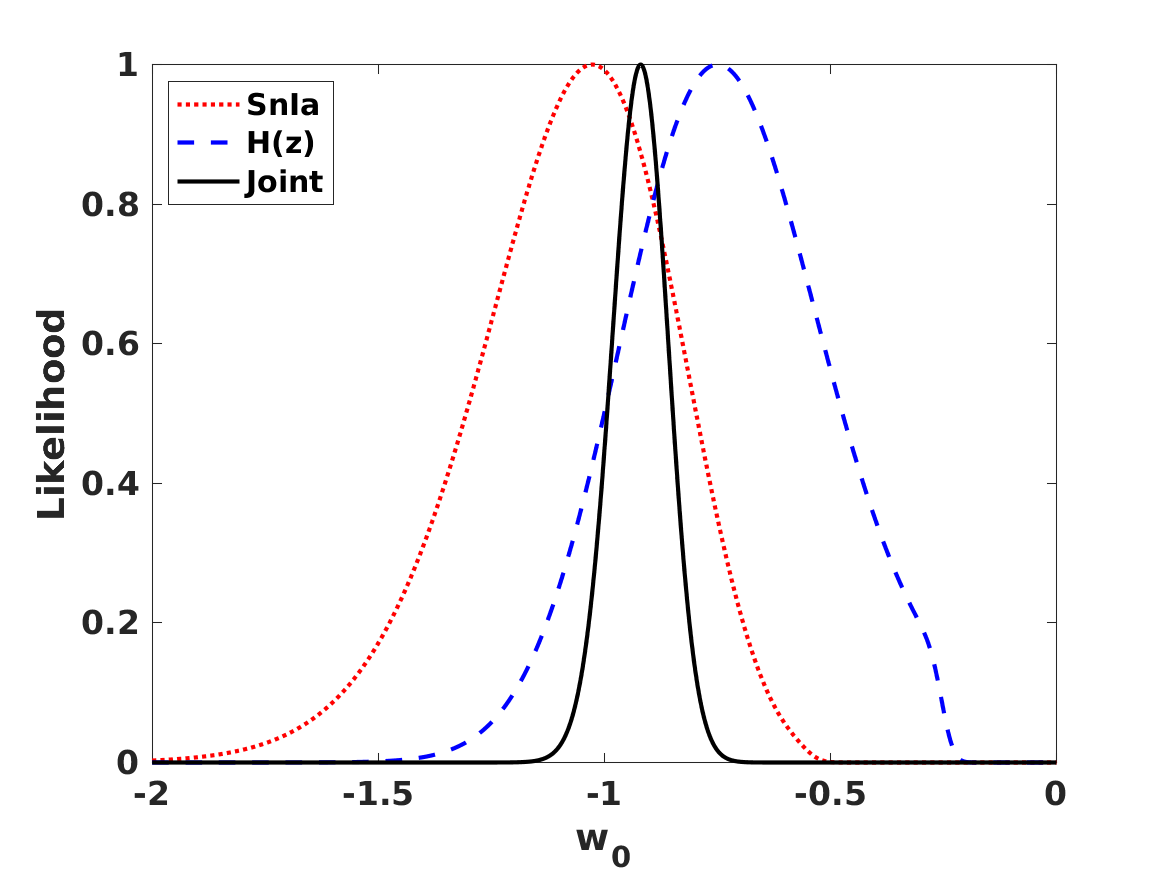}
\includegraphics[width=3.2in,keepaspectratio]{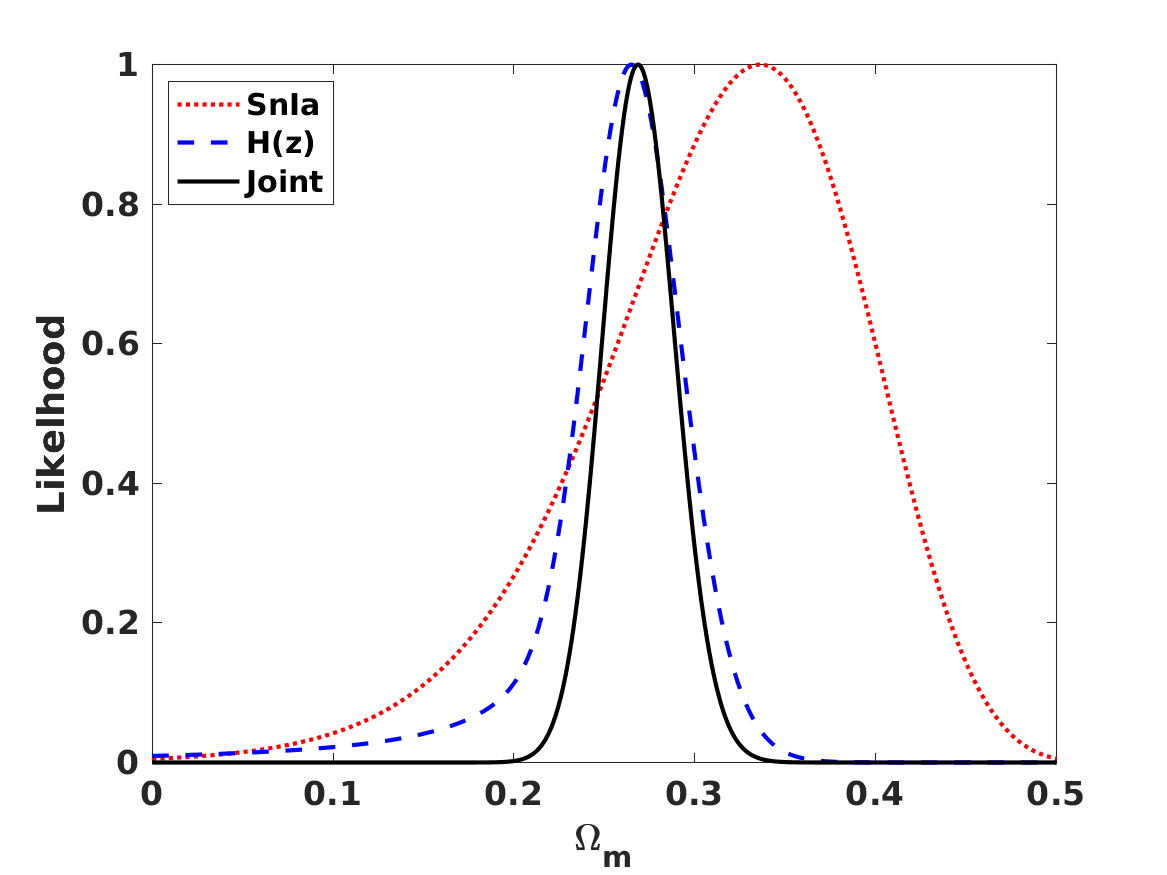}
\end{center}
\caption{\label{fig1}Two-dimensional and one-dimensional (marginalized) constraints for the $w_0$CDM model. In the former case, the one, two and three sigma confidence levels are shown. The red dotted lines show the constraints from the supernova data (denoted 'SnIa' in the legend), the blue dashed lines show the constraints from the Hubble parameter data (denoted 'H(z)'), and the black solid lines show the constraints from the combination of the two data sets (denoted 'Joint').}
\end{figure*}

In what follows we assume a flat Friedmann-Lema\^{\i}tre-Robertson-Walker model, in which case the Friedmann equation has the form
\be
E^2(z)=\Omega_m(1+z)^3+(1-\Omega_m)(1+z)^{3(1+w_0+w_a)}\exp{\left(-\frac{3w_az}{1+z}\right)}\,,
\ee
where the matter parameter is defined as usual, $\Omega_m \equiv \kappa \rho_0 / 3 H_0^2$. This can now be constrained using the aforementioned data.

We start with the simpler case of a constant equation of state parameter, i.e. $w_a=0$. The results of this analysis are shown in Figure \ref{fig1}. We note that the supernova constraints in the $\Omega_m$--$w_0$ plane match those of the left panel of Figure 3 in \cite{Riess}, thus providing a validation test of our analysis code. In this case the one-sigma constraints on the two model parameters from the combined data sets are
\bq
\Omega_m&=&0.27\pm0.02\\
w_0&=&-0.92\pm0.06\,,
\eq
which are compatible with $\Lambda$CDM.

\begin{figure*}
\begin{center}
\includegraphics[width=3.2in,keepaspectratio]{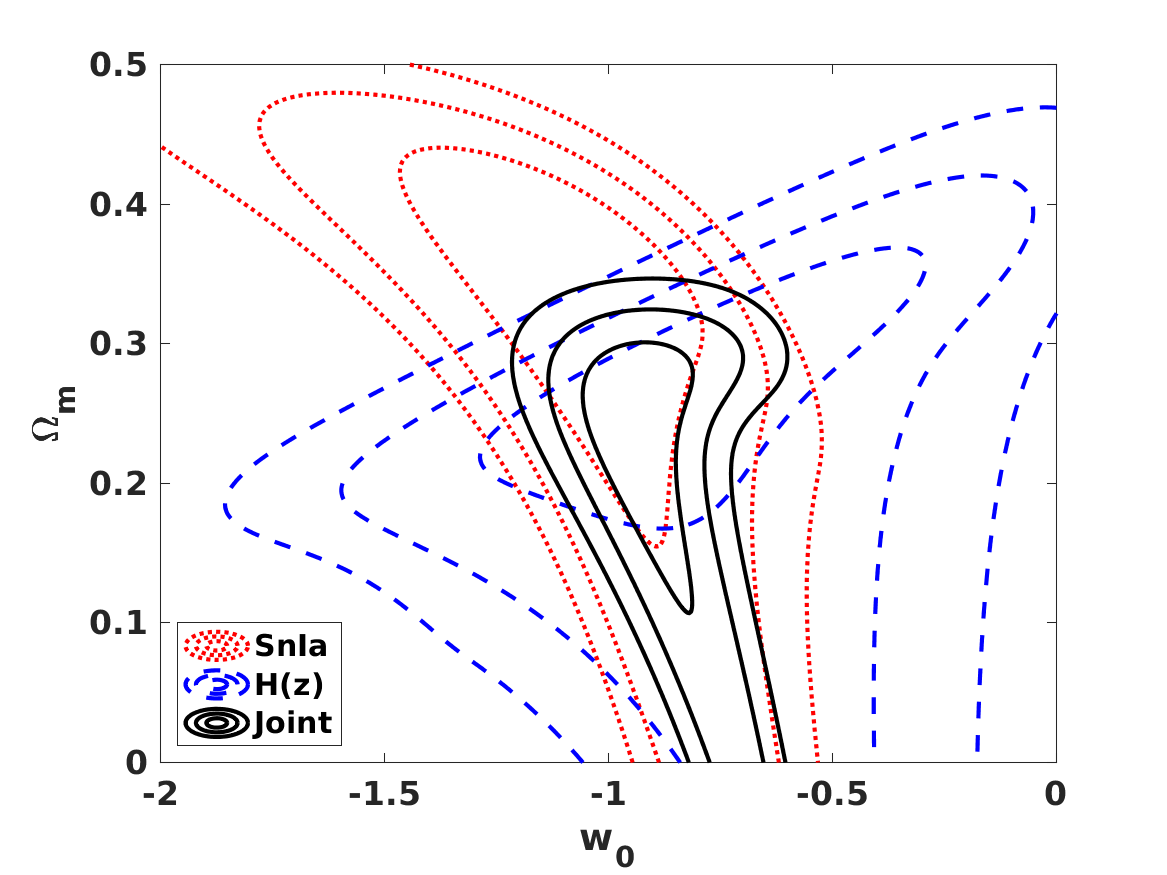}
\includegraphics[width=3.2in,keepaspectratio]{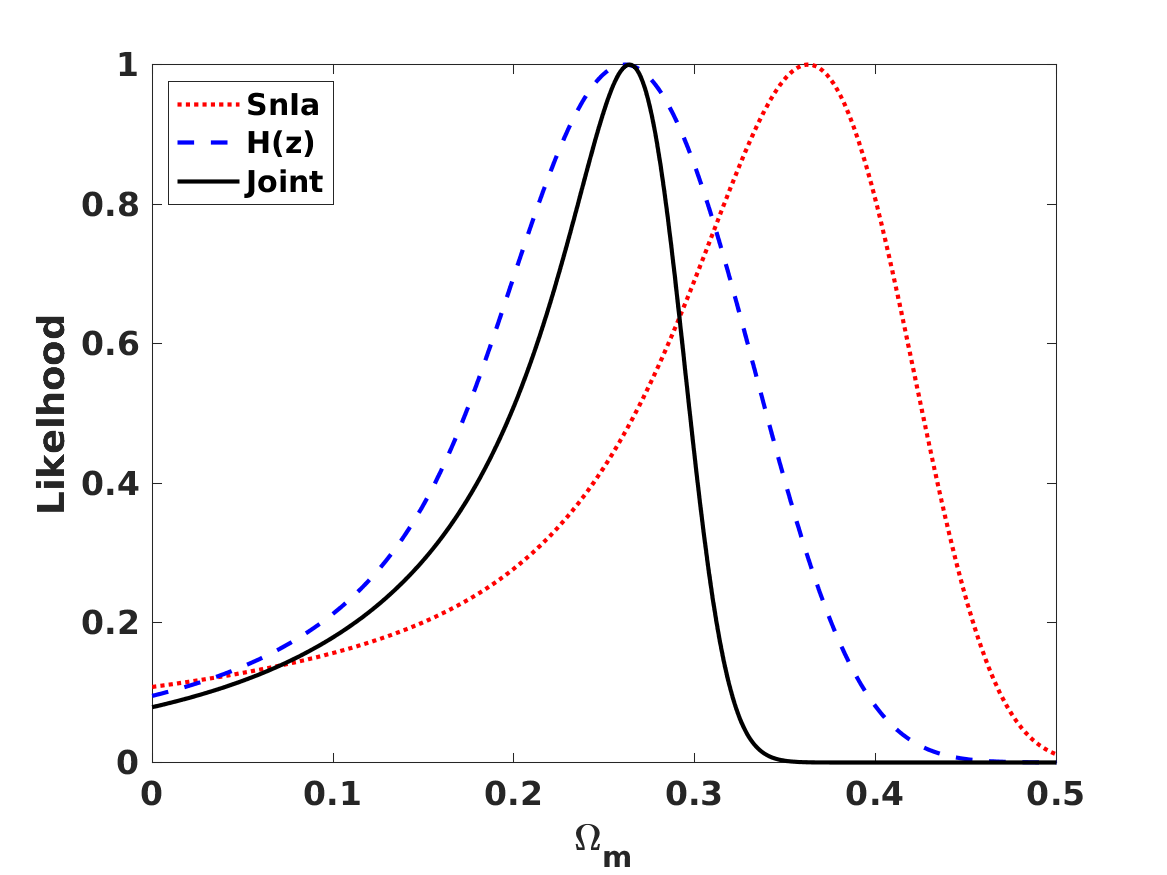}
\includegraphics[width=3.2in,keepaspectratio]{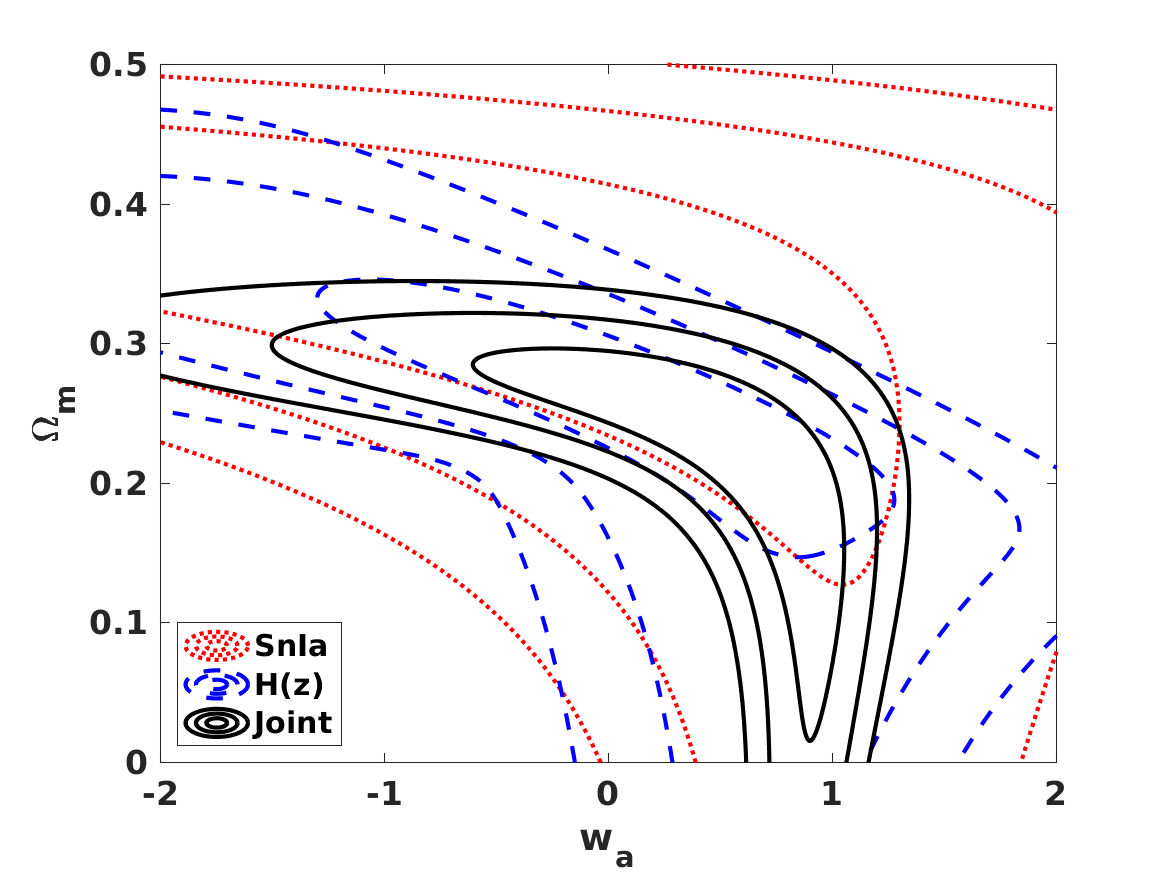}
\includegraphics[width=3.2in,keepaspectratio]{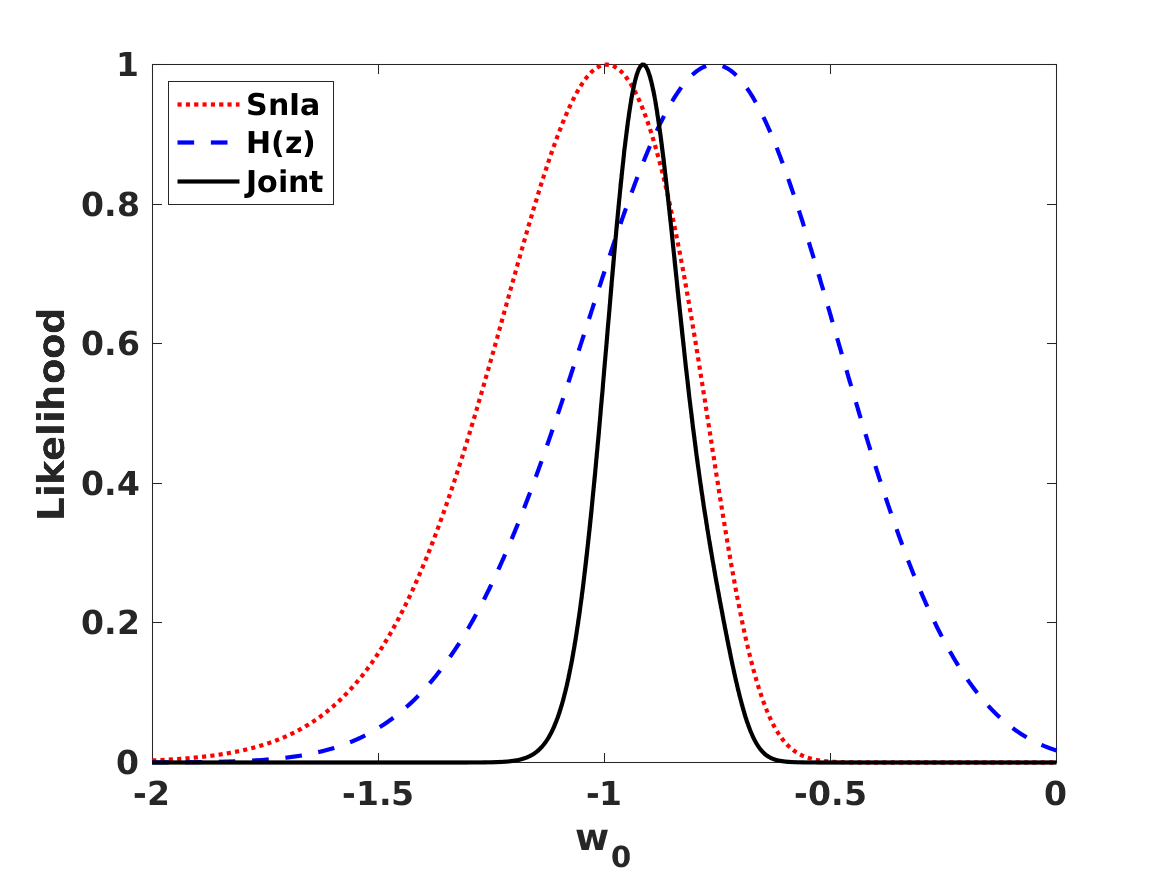}
\includegraphics[width=3.2in,keepaspectratio]{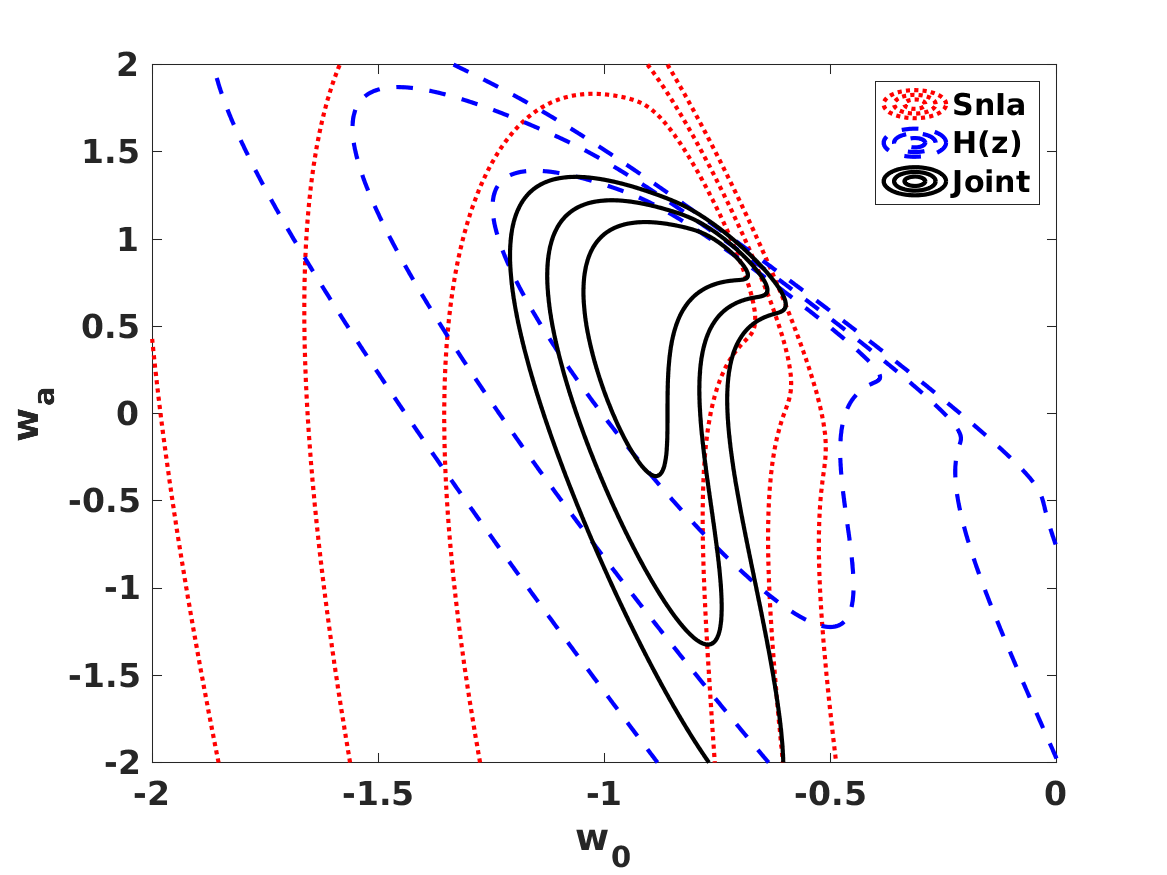}
\includegraphics[width=3.2in,keepaspectratio]{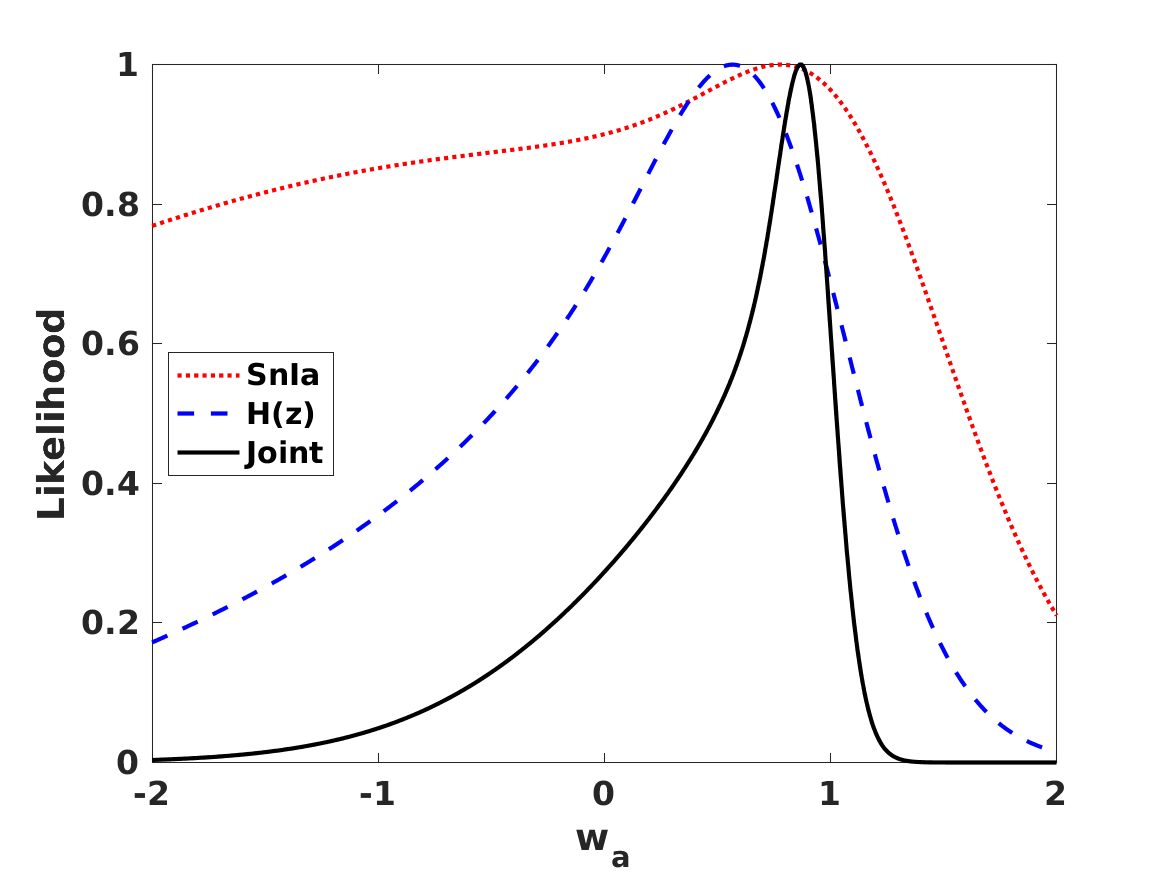}
\end{center}
\caption{\label{fig2}Two-dimensional and one-dimensional constraints (with the remaining parameters marginalized) for the CPL model. In the former case, the one, two and three sigma confidence levels are shown. The red dotted lines show the constraints from the supernova data (denoted 'SnIa' in the legend), the blue dashed lines show the constraints from the Hubble parameter data (denoted 'H(z)'), and the black solid lines show the constraints from the combination of the two data sets (denoted 'Joint').}
\end{figure*}

For the full three-parameter CPL model, the corresponding constraints are shown in Figure \ref{fig2}. Note that in this case a comparison of our supernova constraints in the $w_0$--$w_a$ plane with those of the right panel of Figure 3 in \cite{Riess} is not possible, since the latter include an unspecified Planck prior.

In this case the one-sigma constraints on the three model parameters from the combined data sets are
\bq
\Omega_m&=&0.26^{+0.03}_{-0.05}\\
w_0&=&-0.92^{+0.09}_{-0.08}\\
w_a&=&0.86^{+0.14}_{-0.24};
\eq
the reduced chi-square at the best fit is $\chi^2_\nu\sim0.9$, so the model is slightly overfitting the data (this is mainly driven by the Hubble parameter data). The first two of these constraints are compatible with the values for the $w_0$CDM analysis (with naturally larger uncertainties), but there is a clear preference for a positive slope $w_a>0$. However it is also clear that there are strong degeneracies between the parameters, and moreover the constraints depend on the choice of priors. In the above we used the uniform prior on the matter density $\Omega_m=[0.05,0.5]$, the choice being motivated by the aforementioned validation of our code against the results of \cite{Riess}. Moreover, this broad range will also prove necessary for the model in Sect. \ref{sect4}. If instead one uses the narrower uniform prior $\Omega_m=[0.15,0.45]$, which will be sufficient for the model in Sect. \ref{sect3}, we find
\bq
\Omega_m&=&0.26^{+0.03}_{-0.05}\\
w_0&=&-0.92^{+0.07}_{-0.08}\\
w_a&=&0.74^{+0.21}_{-0.48};
\eq
in other words, there is no impact on the matter density and $w_0$, but there is a significant impact on $w_a$. Breaking these degeneracies requires additional data, for example from cosmic microwave background observations.

In any case, our main goal in the above analysis is to set up a benchmark for the constraining power of these data sets, against which to compare the constraints on the alternative models to be discussed in what follows.


\section{Generalized couplings: the Feng-Carloni model}
\label{sect3}

The precise nature of the coupling between matter and the metric in the Einstein equations is one of the most questionable assumptions of the theory. One may therefore explore the possibility that this coupling is nontrivial. In Feng and Carloni's generalized coupling model \cite{Feng} one assumes a coupling of the form
\begin{equation}
    G_{\mu \nu} = \chi_{\mu \nu}^{\alpha \beta} T_{\alpha \beta}
\end{equation}
where $\chi_{\mu \nu}^{\alpha \beta}$ is a nonsingular fourth-order tensor, subject to the constraint that in vacuum $\chi_{\mu \nu}^{\alpha \beta} = \kappa \delta_\mu^\alpha \delta_\nu^\beta$, where for future convenience we have defined $\kappa=8\pi G$. This ensures that the theory is equivalent to General Relativity in vacuum, but still allows for a different behaviour within a matter distribution.

For concreteness, in \cite{Feng} the authors assume that the gravitational metric is coupled to matter through an auxiliary rank-2 tensor
\be
\chi_{\mu \nu}^{\alpha \beta} = \phi(A) A_\mu^\alpha A_\nu^\beta\,,
\ee
where $A_\mu^\alpha$ is non-dynamical (no derivatives of it appear in the action) and reverts to $\delta_\mu^\alpha$ in vacuum (in accordance with the condition in the previous paragraph) and $\phi$ is a scalar function of $A_\mu^\alpha$.

Under these assumptions, \cite{Feng} obtain the following Friedmann and Raychaudhuri equations
\begin{equation} \label{friedmann_2}
    3q \left( H^2+\frac{k}{a^2} \right)=\frac{256\kappa(1-pq)^3(q\rho + 1)^2}{[4+q(\rho-3p)]^4} + q\Lambda - \kappa
\end{equation}
\bq \label{raychaudhuri_2}
    6q(\dot{H} + H^2) &=& \frac{256\kappa(1-pq)^3(q\rho + 1)[2-q(\rho+3p)]}{[4+q(\rho-3p)]^4} \nonumber \\
    && + 2(q\Lambda - \kappa)\,,
\eq
where $a$ is the scale factor, $k$ (not to be confused with $\kappa$) is the usual curvature parameter, $\Lambda$ is the cosmological constant, $q$ is a model-specific parameter (to be further discussed shortly) and $p$ is the pressure of a fluid that is assumed to be barotropic, with an equation of state $p=w\rho$, where $w$ is a constant equation of state parameter. In what follows we will further assume a flat universe (that is $k=0$). The corresponding continuity equation takes the form
\begin{equation} \label{continuity_2}
    \dot{\rho} = -\frac{3H\rho(w+1)[q^2\rho^2w(3w-1)+q\rho (1-7w) + 4]}{q^2 \rho^2w(3w-1) - q\rho(3w^2+13w+2)+4}\,.
\end{equation}

The parameter $q$ is defined as
\be
q=\frac{\kappa}{\lambda}\,,
\ee
where $\lambda$ is interpreted in \cite{Feng} as being akin to the vacuum energy density generated by matter fields (and thus also an energy scale at which a field theory description of the model breaks down). Note that this implies that the model contains two kinds of vacuum energy. The authors argue that $q<0$ if the contribution from fermions is the dominant one, and that $q<0$ is necessary for the dynamical stability of the theory and also enables inflationary-type solutions at early times. On the other hand, this is effectively a bimetric theory, with different propagation speeds (in matter) for gravitons and photons, and $q<0$ would lead to superluminal gravitational wave propagation, while the $q>0$ case is well-behaved. Also, note that large positive values of $q$ would lead to a singularity in the continuity equation, while large negative values of $q$ can lead to a singularity in the Friedmann equation. In other words, it's clear that the parameter $q$ is constrained to be small in absolute value. Therefore, as previously mentioned, in what follows we take the model as a phenomenological one and treat $q$ (or a dimensionless version thereof) as a free parameter to be constrained by the data.

\begin{figure*}
\begin{center}
\includegraphics[width=3.2in,keepaspectratio]{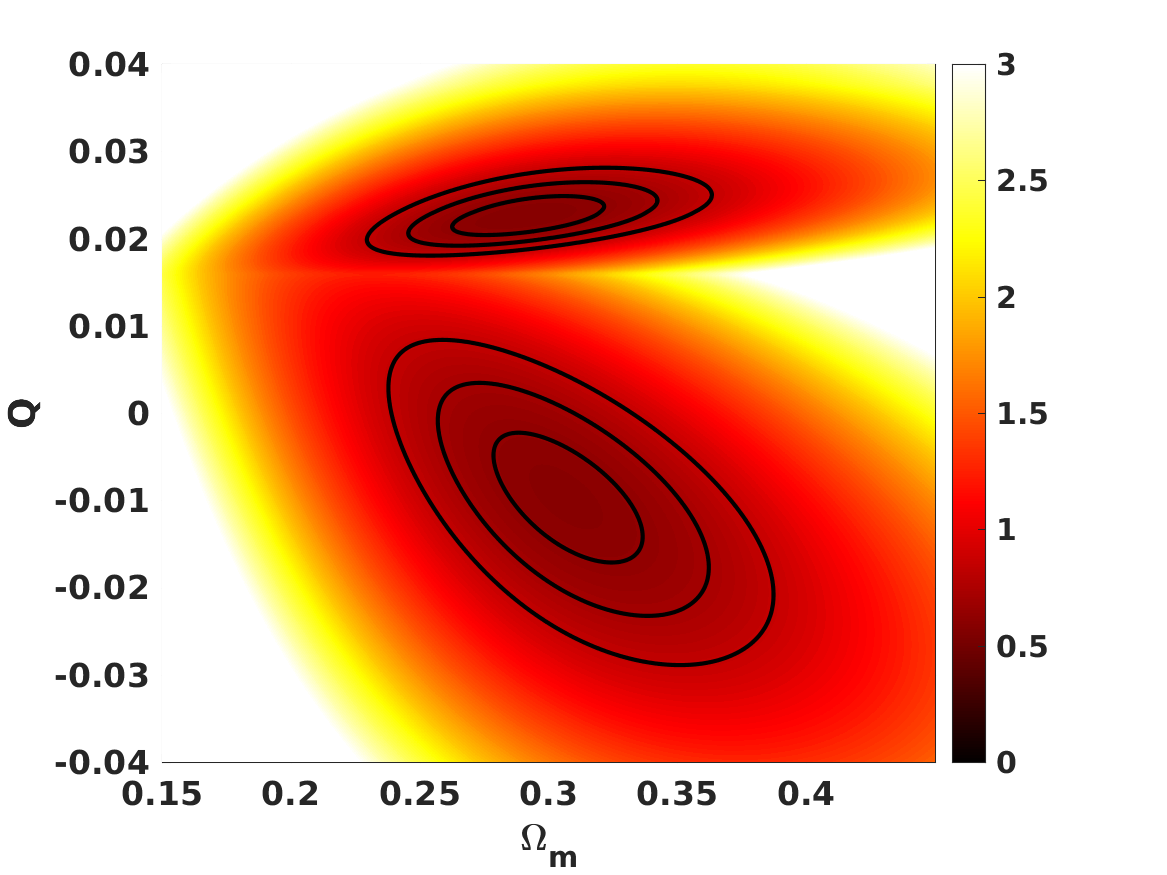}\\
\includegraphics[width=3.2in,keepaspectratio]{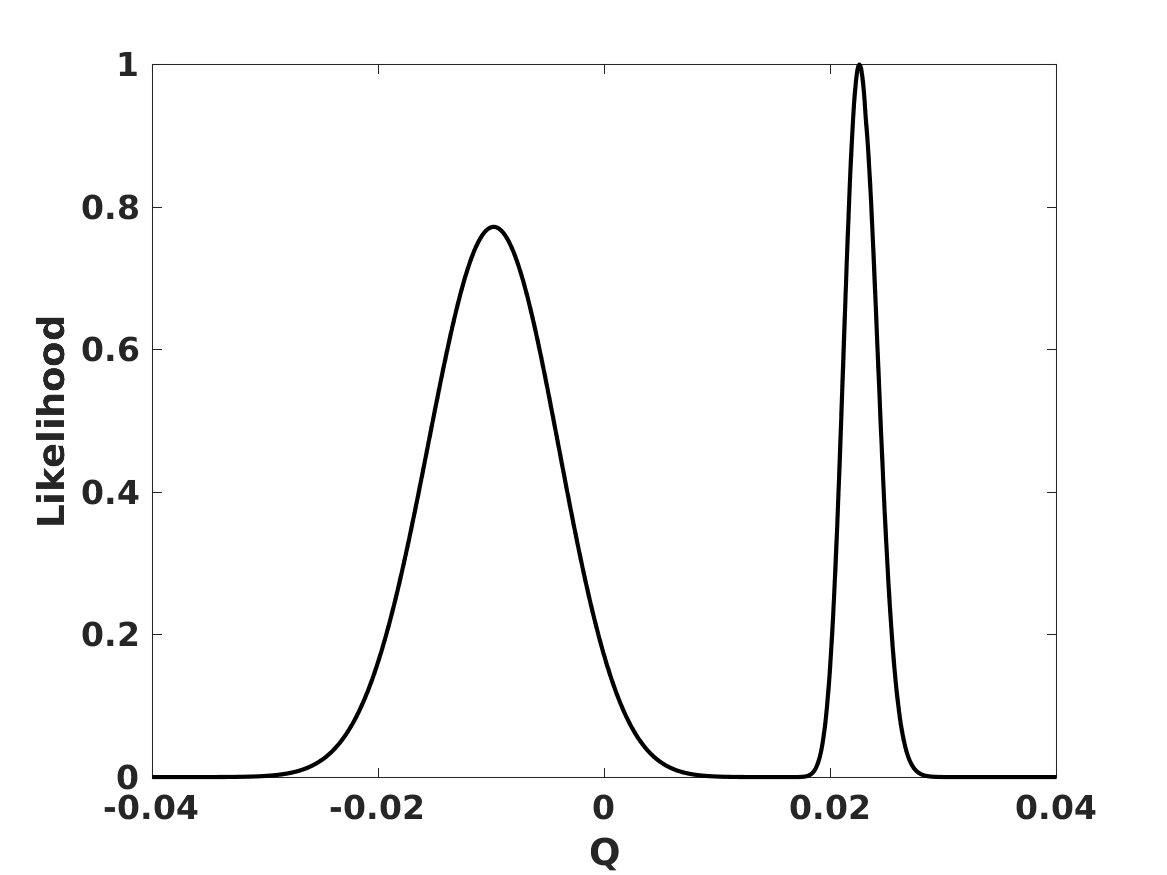}
\includegraphics[width=3.2in,keepaspectratio]{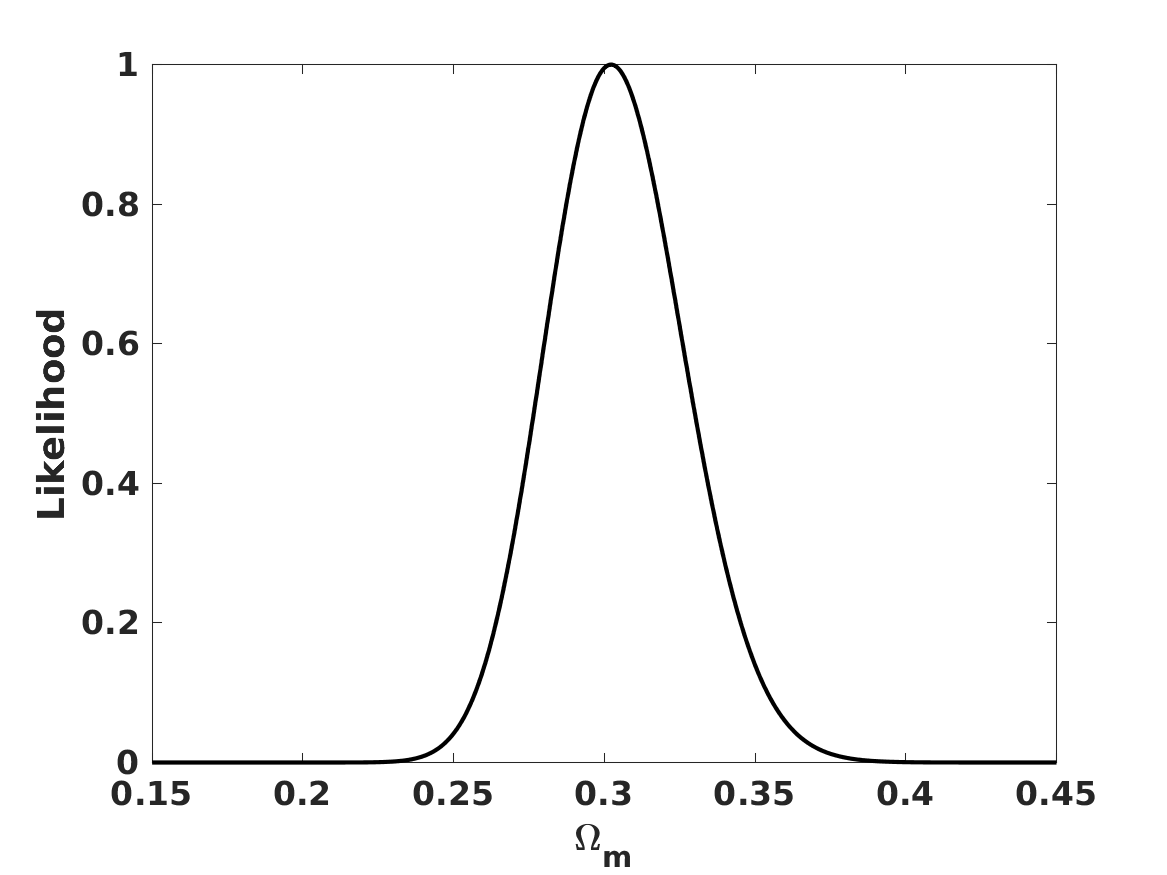}
\end{center}
\caption{\label{fig3}Two-dimensional and one-dimensional constraints (with the remaining parameter marginalized) for the $w=0$ generalized coupling model, for the combination of supernova and Hubble parameter data. In the two-dimensional case, the $\Delta\chi^2=2.3$, $\Delta\chi^2=6.17$ and $\Delta\chi^2=11.8$ confidence levels are shown in black lines, and the color map depicts the reduced chi-square at each point in the parameter space, with points with $\chi^2_\nu>3$ shown in white.}
\end{figure*}

As usual, only two of the above three equations are independent, and we choose to work with the Friedmann and continuity equations. Moreover, we also define
\be
r\equiv\frac{\rho}{\rho_0}
\ee
\be
Q\equiv q\rho_0\,,
\ee
where $\rho_0$ is the present-day critical density. We thus re-write the two equations in following form
\bq
E^2(z) &=& \frac{f_1(Qr,w)}{\left[1+\frac{1-3w}{4}Qr\right]^4}r\Omega_m + \Omega_\Lambda \\
\frac{dr}{dz} &=& 3\frac{r(1+w)}{1+z}f_2(Qr,r)\,,
\eq
where
\bq
f_1(Qr,w)&=&1+c_1(Qr)+c_2(Qr)^2+c_3(Qr)^3-w^3(Qr)^4\\
c_1&=&-\frac{1}{8}(3w^2+30w-5)\\
c_2&=&\frac{1}{16}(11w^3+69w^2-39w-1)\\
c_3&=&-\frac{1}{256}(8w^4+404w^3-714w^2-12w+1)\\
f_2(Qr,w)&=&\frac{Q^2r^2w(3w-1)+Q r (1-7w) + 4}{Q r^2w(3w-1) - Q r(3w^2+13w+2)+4}\,.
\eq
This equivalent formulation is algebraically lengthier but numerically more robust, being well-behaved in the $Q\longrightarrow0$ limit. We have kept the usual matter and dark energy densities, $\Omega_m$ and $\Omega_\Lambda\equiv\Lambda/ 3 H_0^2$, separately in the Friedmann equation to make it more explicit that the standard $\Lambda$CDM model is recovered (at least at the background level) in the $Q\longrightarrow0$ limit, but as previously mentioned we will restrict our analysis of this model to flat universes, so $\Omega_m$ and $\Omega_\Lambda$ are not independent.

\begin{figure*}
\begin{center}
\includegraphics[width=3.2in,keepaspectratio]{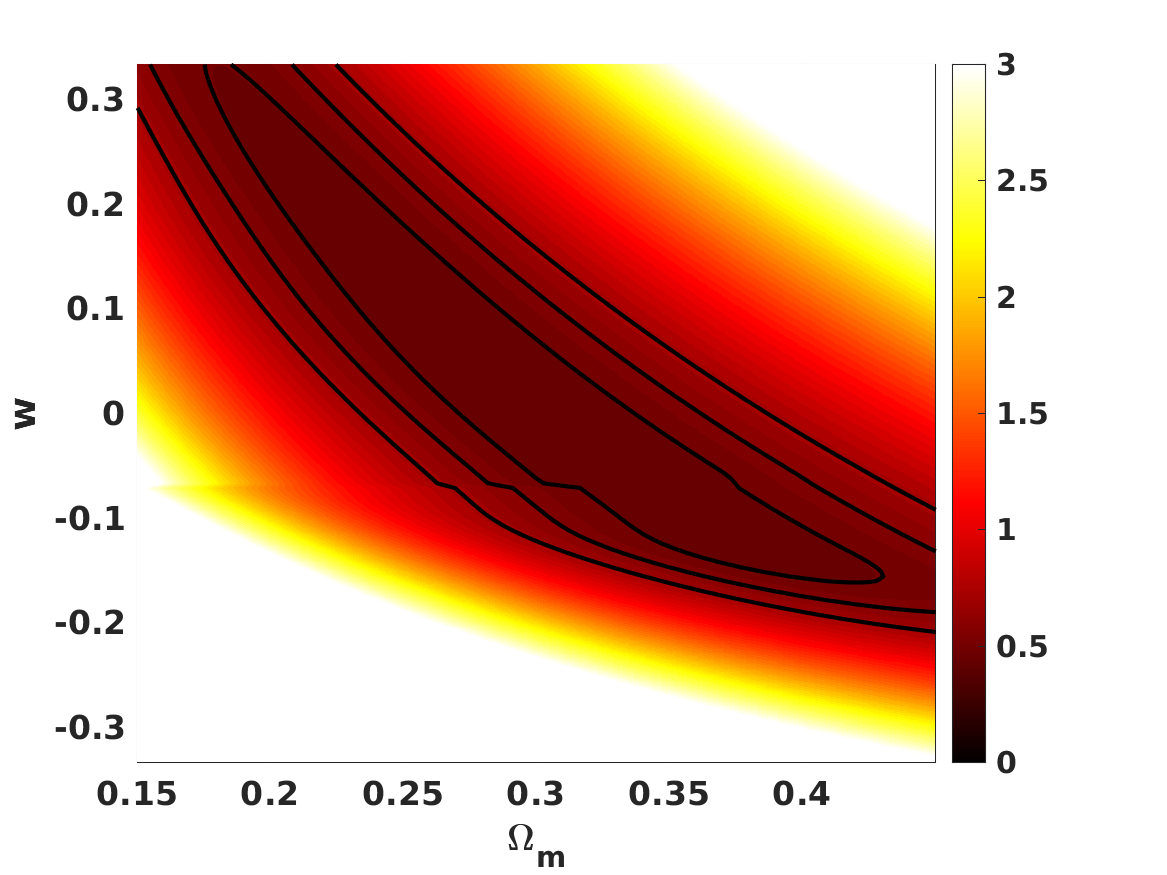}
\includegraphics[width=3.2in,keepaspectratio]{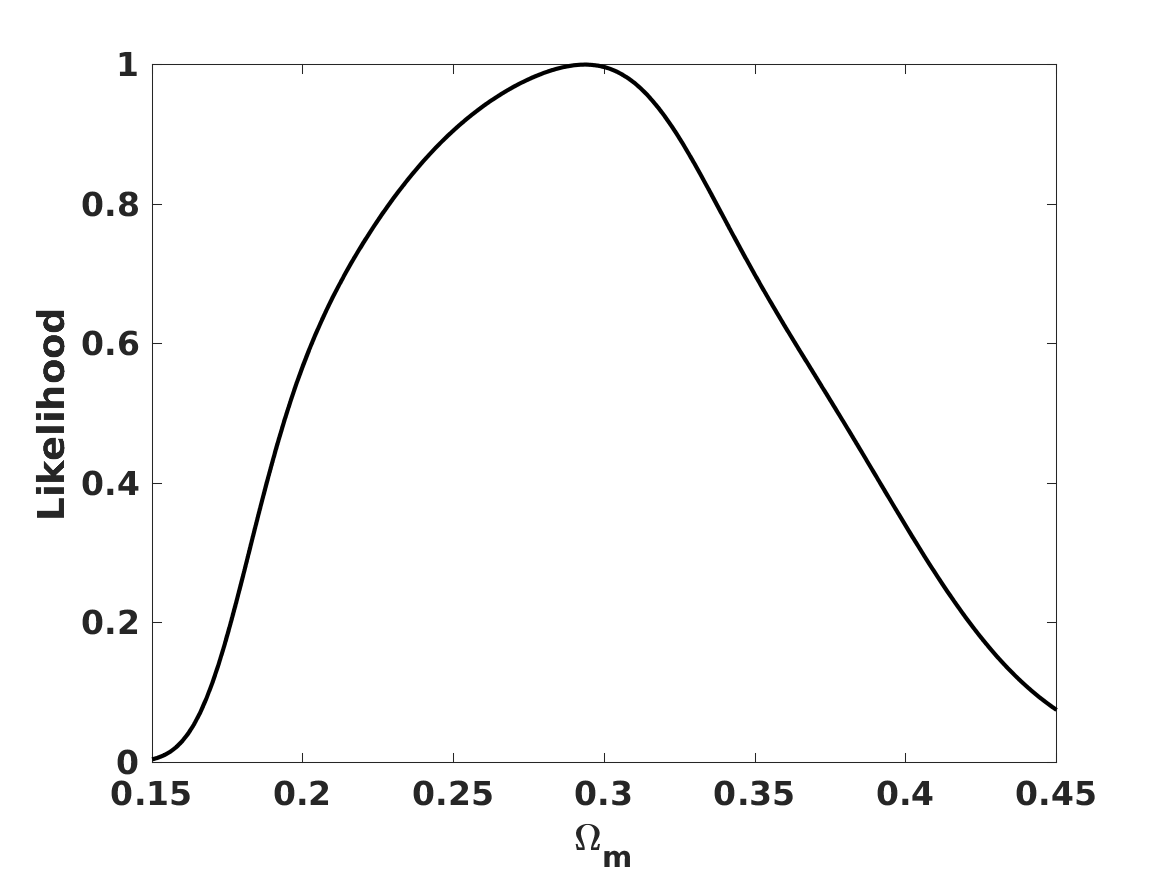}
\includegraphics[width=3.2in,keepaspectratio]{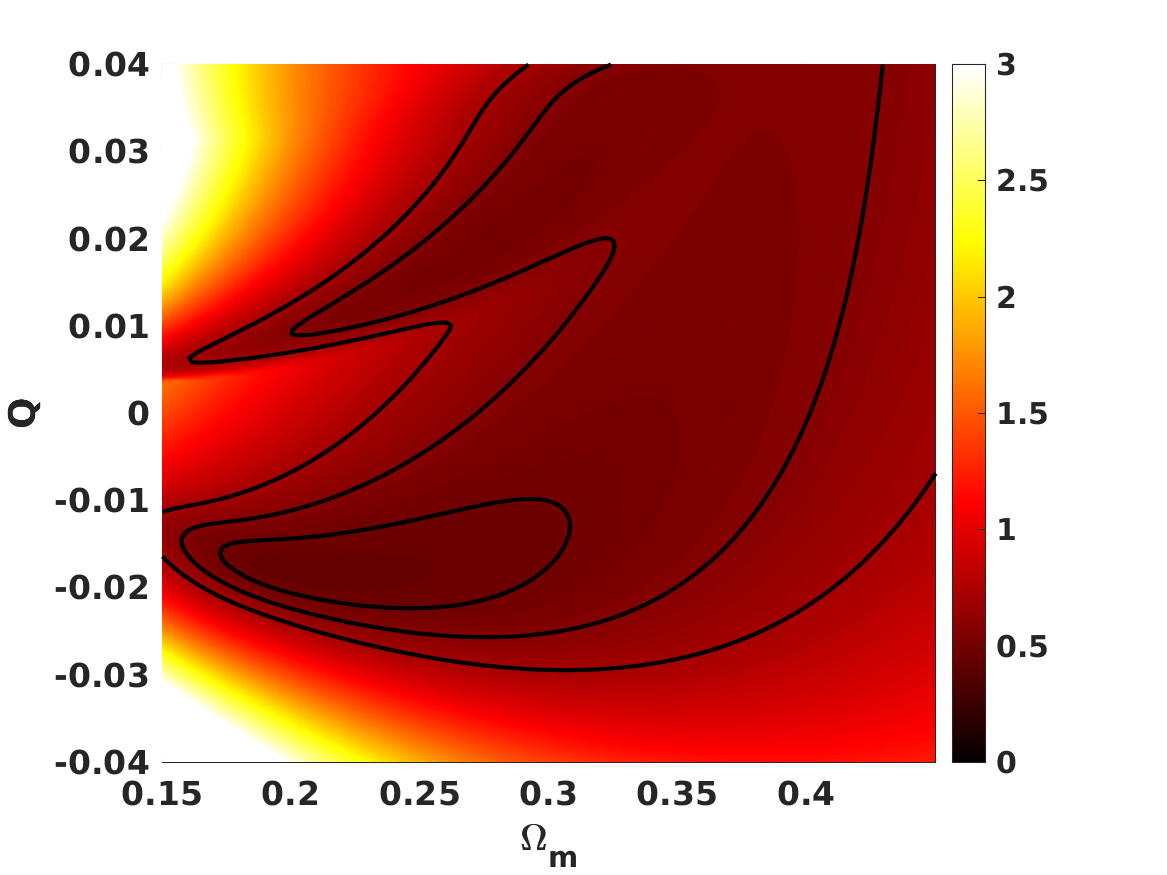}
\includegraphics[width=3.2in,keepaspectratio]{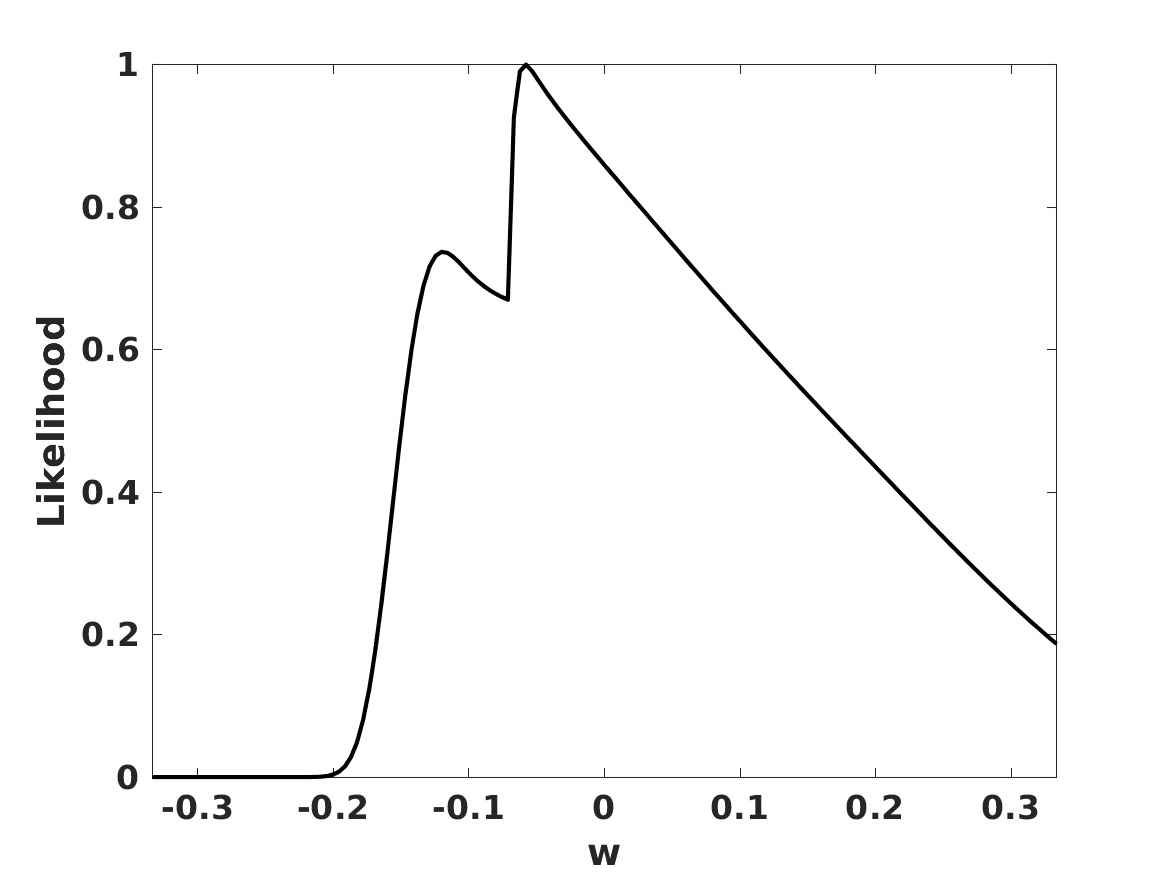}
\includegraphics[width=3.2in,keepaspectratio]{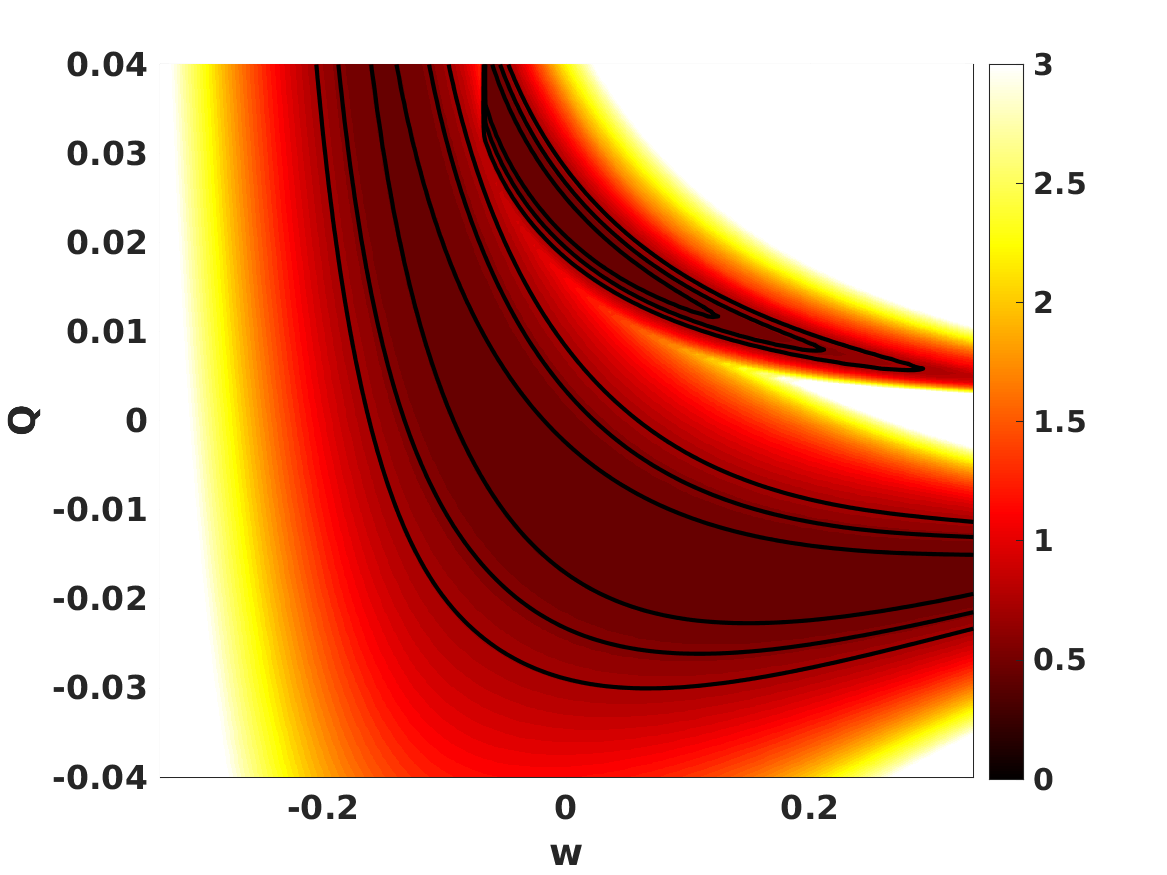}
\includegraphics[width=3.2in,keepaspectratio]{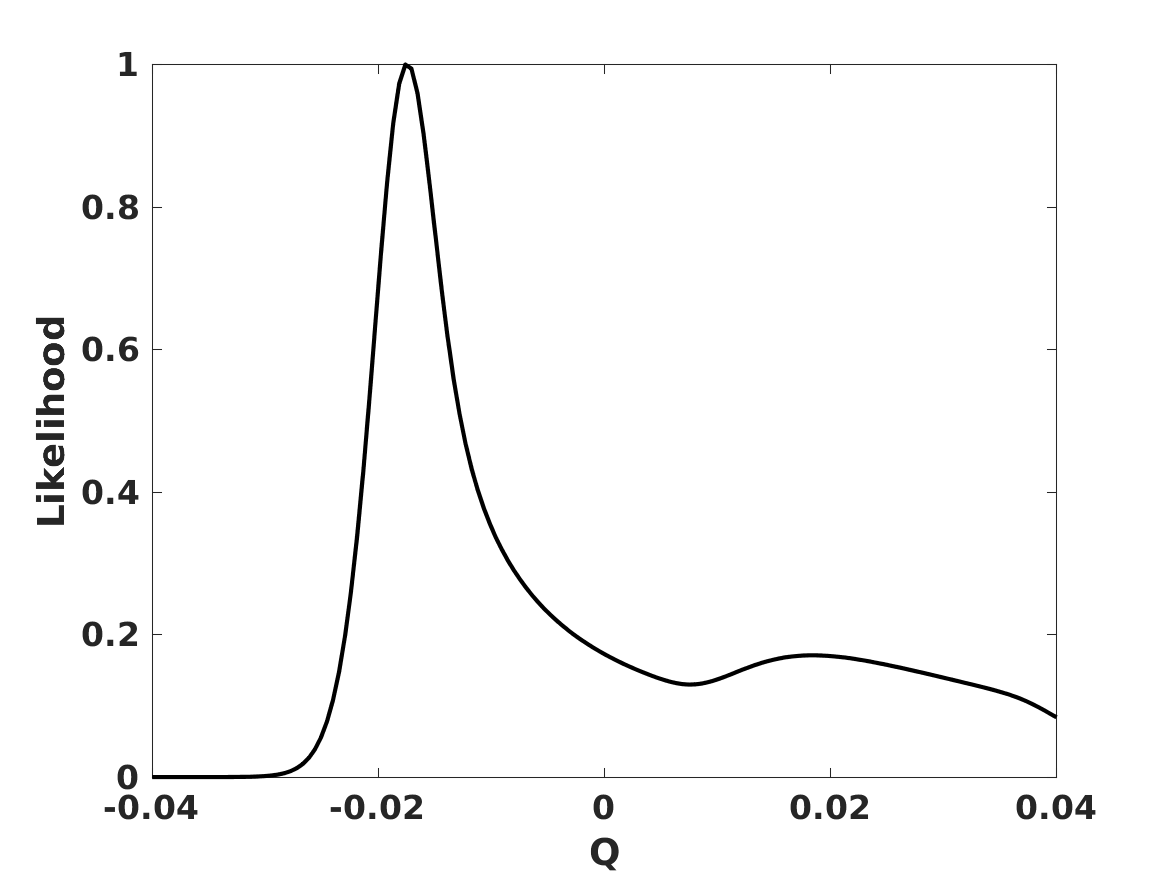}
\end{center}
\caption{\label{fig4}Two-dimensional and one-dimensional constraints (with the remaining parameters marginalized) for the full generalized coupling model, with $w$ allowed to vary, for the combination of supernova and Hubble parameter data. In the two-dimensional case, the $\Delta\chi^2=2.3$, $\Delta\chi^2=6.17$ and $\Delta\chi^2=11.8$ confidence levels are shown in black lines, and the color map depicts the reduced chi-square at each point in the parameter space, with points with $\chi^2_\nu>3$ shown in white.}
\end{figure*}

The model has three free parameters: the aforementioned $Q$, the matter density $\Omega_m$ and the value of the equation of state parameter $w$. As a simple illustration of the impact of the parameter $Q$ on the matter density, we note that linearizing the Friedmann and continuity equation one finds the approximate solution for the redshift dependence of the (dimensionless) matter density 
\be
r(z)\approx (1+z)^{3(1+w)}\left(1+\frac{3}{4}(1+w)^2Q\left[(1+z)^{3(1+w)}-1\right]\right)\,.
\ee
On the other hand, for $w=0$ one finds the exact but implicit solution
\be
\frac{r}{(1+Qr/4)^3}=\frac{(1+z)^3}{(1+Q/4)^3}\,.
\ee
We note that this implies that the matter density does not follow the standard behaviour $\rho\propto (1+z)^3$, and two relevant points stem from this. The first is that some of the measurements in the compilation of Hubble parameter measurements that we use \cite{Farooq} are based on the horizon sound scale, which is inferred assuming a standard matter content. On the other hand, one may expect that this non-standard behaviour of the matter density will itself lead to a tight constraint on the angular size of the sound horizon \cite{Amendola}. Thus our constraints in this case are only approximate, but we expect them to be conservative ones, since the latter constraint should be a strong one.  That said we also note that the highly non-linear dependence on the parameter $Q$ raises the possibility that several choices of the parameter may simultaneously fit the data---an expectation that will be partially confirmed presently.

Note that since the model effectively has two types of vacuum energy, viz. the one generated by matter fields as well as the usual cosmological constant, one may wonder if the former is sufficient to yield an accelerating universe without invoking the latter. However, it is simple to show that this can't be the case. If one sets $\Omega_\Lambda=0$ in the Friedmann equation (taking also the matter equation of state parameter $w=0$ for simplicity), the parameters $\Omega_m$ and $Q$ are no longer independent, being related by
\be
\Omega_m=\frac{32(1+Q/4)^4}{(1-Q/8)(Q^2+24Q+32)}\,;
\ee
the requirement $0\le\Omega_m\le1$ corresponds to $-1\le Q\le0$, with $\Omega_m=1$ ensuing both for $Q=0$ (corresponding to the Einstein-de Sitter case) and for $Q=-1$. In this range of $Q$ the minimum density is $\Omega_m\sim0.865$, corresponding to $Q\sim-0.62$; clearly such high matter density universes would be incompatible with observations. Thus in what follows we treat this model as a phenomenological extension of $\Lambda$CDM, with the vacuum energy density of matter fields, $Q$, being an additional model parameter which we now constrain.

We begin again by considering the simpler case where the matter equation of state parameter has the standard value, $w=0$, leaving only two free parameters, and agnostically allowing both positive and negative values of the model parameter $Q$. The results of this analysis are depicted in figure \ref{fig3}: while non-zero values of $Q$ are preferred, the standard value is not significantly excluded. We note the existence of two branches of the solution, one with $Q>0$ and the other with $Q<0$, with the former branch being slightly preferred. If we restrict the analysis to the range $Q\le0$, the one-sigma constraints on the two model parameters are
\bq
\Omega_{m-} &=& 0.31\pm 0.02\\
Q_-&=&-0.010\pm0.006\,;
\eq
conversely, if we restrict the analysis to the range $Q\ge0$ we find
\bq
\Omega_{m+} &=& 0.29\pm 0.02\\
Q_+&=&0.023\pm0.003\,.
\eq
In all cases the reduced chi-square at the best fit is $\chi^2_\nu\sim0.6$, so the model is clearly overfitting the data. This, together with the presence of the two branches of the solution, implies that there is no strong evidence for a non-zero $Q$.

We now consider the general case, allowing $w$ to become a further free parameter. The results are depicted in figure \ref{fig4}, and the corresponding one-sigma constraints on the parameters are
\bq
    \Omega_m &=& 0.29^{-0.09}_{+0.07}\\
    Q &=& -0.018^{+0.005}_{-0.004}\\
    w &=& -0.06^{+0.17}_{-0.08}\,.
\eq
Although the constraints on the matter density are now significantly weaker, two branches of the solution are still manifest, as are the degeneracies between the model parameters. In this case the negative branch is the preferred one. However, it also has to be said that the matter equation of state parameter is already more tightly constrained than this. Recent analyses \cite{Thomas,Tutusaus} constrain it, conservatively, to $|w|<0.003$. Using this as a Gaussian prior and repeating the analysis, we recover the constraints on $Q$ and $\Omega_m$ previously discussed for the $w=0$ case, while the posterior for $w$ itself simply recovers the prior.

\section{Scale invariance: the Maeder model}
\label{sect4}

The recently proposed scale invariant model \cite{Maeder} draws heavily on previous work on scale-covariant theories by Canuto \textit{et al.} \cite{Canuto1,Canuto2}. Although it is well known that the effects of scale invariance are expected to disappear upon the presence of matter, the assumption underlying scale invariant models is that at large (i.e., cosmological) scales empty space should still be scale invariant. This assumption ultimately leads to a bimetric theory, with a function $\lambda$ (not to be confused with the parameter introduced in the previous section) playing the role of a scale transformation factor relating the ordinary matter frame to another frame which one assumes to still be scale invariant.

In this case, and with the further assumption of a homogeneous and isotropic universe, the Friedmann, Raychaudhuri, and continuity equations have the following form \cite{Canuto1,Canuto2}
\bq
\left(\frac{\dot a}{a}+\frac{\dot\lambda}{\lambda}\right)^2+\frac{k}{a^2}&=&\frac{1}{3}(\kappa\rho+\Lambda\lambda^2)\\
\frac{\ddot a}{a}+\frac{\ddot \lambda}{\lambda}+\frac{\dot \lambda}{\lambda}\frac{\dot a}{a}-\frac{{\dot\lambda}^2}{\lambda^2}&=&-\frac{\kappa}{6}(\rho+3p-2\Lambda\lambda^2)\\
{\dot\rho}+3(\rho+p)\frac{\dot a}{a}&=&-(\rho+3p)\frac{\dot \lambda}{\lambda}\,,
\eq
which trivially reproduce the standard equations for $\lambda=1$. Note that for a homogeneous and isotropic model $\lambda$ depends only on time, as does the scale factor.

To this, the recent work of Maeder adds a further assumption, \textit{viz.} that the Minkowski metric is a solution of these Einstein equations. This leads to the following consistency conditions \cite{Maeder}
\bq
3\frac{{\dot\lambda}^2}{\lambda^2}&=&\Lambda\lambda^2\\
2\frac{\ddot \lambda}{\lambda}-\frac{{\dot\lambda}^2}{\lambda^2}&=&\Lambda\lambda^2\,,
\eq
which can be used to simplify the Friedmann and Raychaudhuri equations. Moreover, the two consistency conditions imply
\be
\lambda=\sqrt{\frac{3}{\Lambda}}\frac{1}{t}\,;
\ee
note that this assumes $c=1$. Just as in the previous section, we will assume constant equations of state, $p=w\rho$. Together with the solution for $\lambda$, the continuity equation yields
 \be
 \rho\propto (1+z)^{3(1+w)}t^{1+3w}\,;
 \ee
in particular, for a cosmological constant ($w=-1$) this becomes $\rho\propto t^{-2}$. In other words, the Maeder assumption effectively leads to a model with a time-dependent cosmological constant, but no parametric $\Lambda$CDM limit. In \cite{Maeder} the author claims, from a simple qualitative comparison, that with the choice $\Omega_m=0.3$ the model is in good agreement with Hubble parameter data. In what follows we assess this claim with a more thorough statistical analysis.

With the aforementioned assumptions, the Friedmann equation for the Maeder model can be written
\begin{equation}
E^2(z,x)=\Omega_m(1+z)^{3(1+w)}x^{1+3w}+\Omega_k(1+z)^2+\frac{\Omega_\lambda}{x}E(z,x)\,,
\label{scalef}
\end{equation}
where the matter parameter $\Omega_m$ has the standard definition, $\Omega_k=-k/(a_0H_0)^2$, and we have also defined an effective dark energy parameter $\Omega_\lambda=2/(t_0H_0)$ and a dimensionless time $x=t/t_0$, with $t_0$ being the current age of the universe.

Actually this Friedmann equation can be re-written in the simpler form
\bq
E(z,x)&=&\frac{\Omega_\lambda}{2x}\left[]1+\sqrt{1+M(z,x)}\right]\\
M(z,x)&=&\frac{4}{\Omega_\lambda^2}\left[\Omega_m(1+z)^{3(1+w)}x^{3(1+w)}+\Omega_k(1+z)^2x^2\right]\,,
\eq
with the relation between redshift and (dimensionless) time being given by
\be
\frac{dx}{dz}=- \frac{x}{1+z}\times\frac{1}{1+\sqrt{1+M(z,x)}}\,,
\ee
with the initial condition $x=1$ at $z=0$.

\begin{figure*}
\begin{center}
\includegraphics[width=3.2in,keepaspectratio]{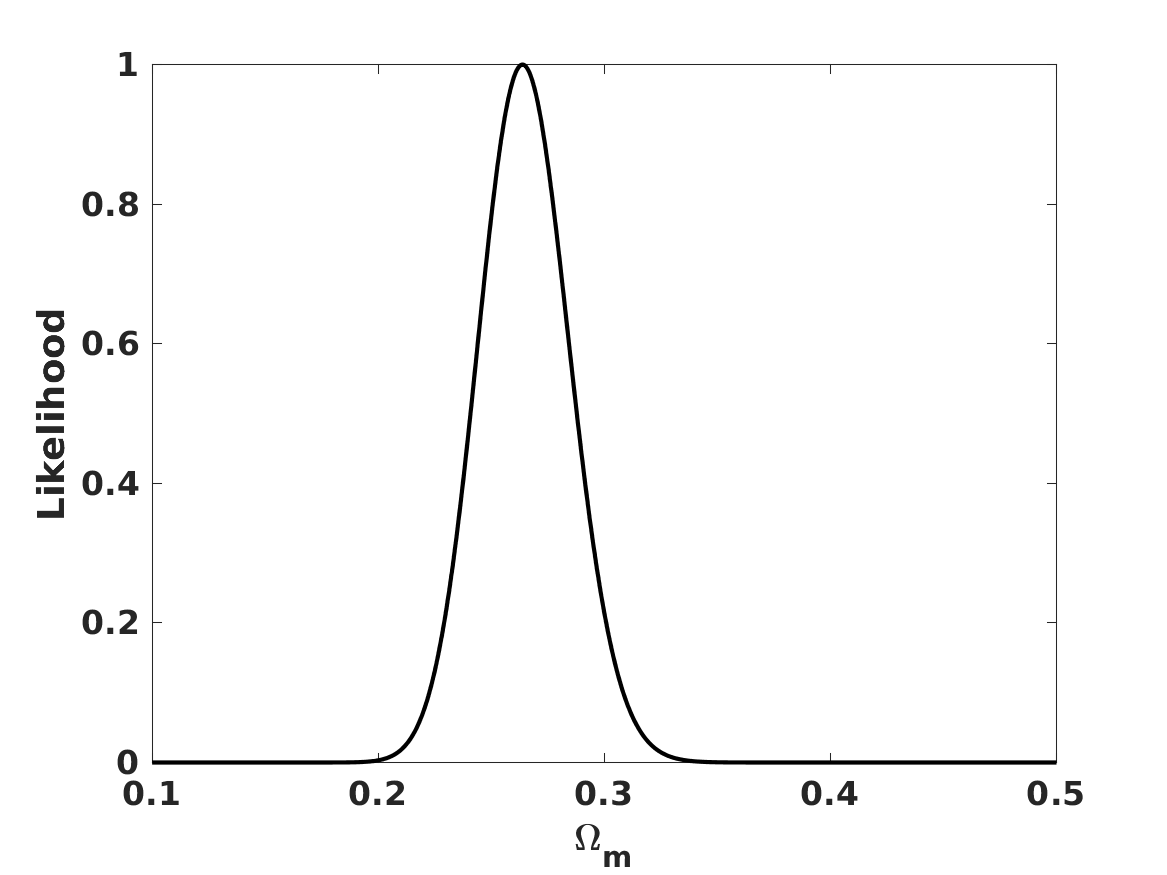}
\includegraphics[width=3.2in,keepaspectratio]{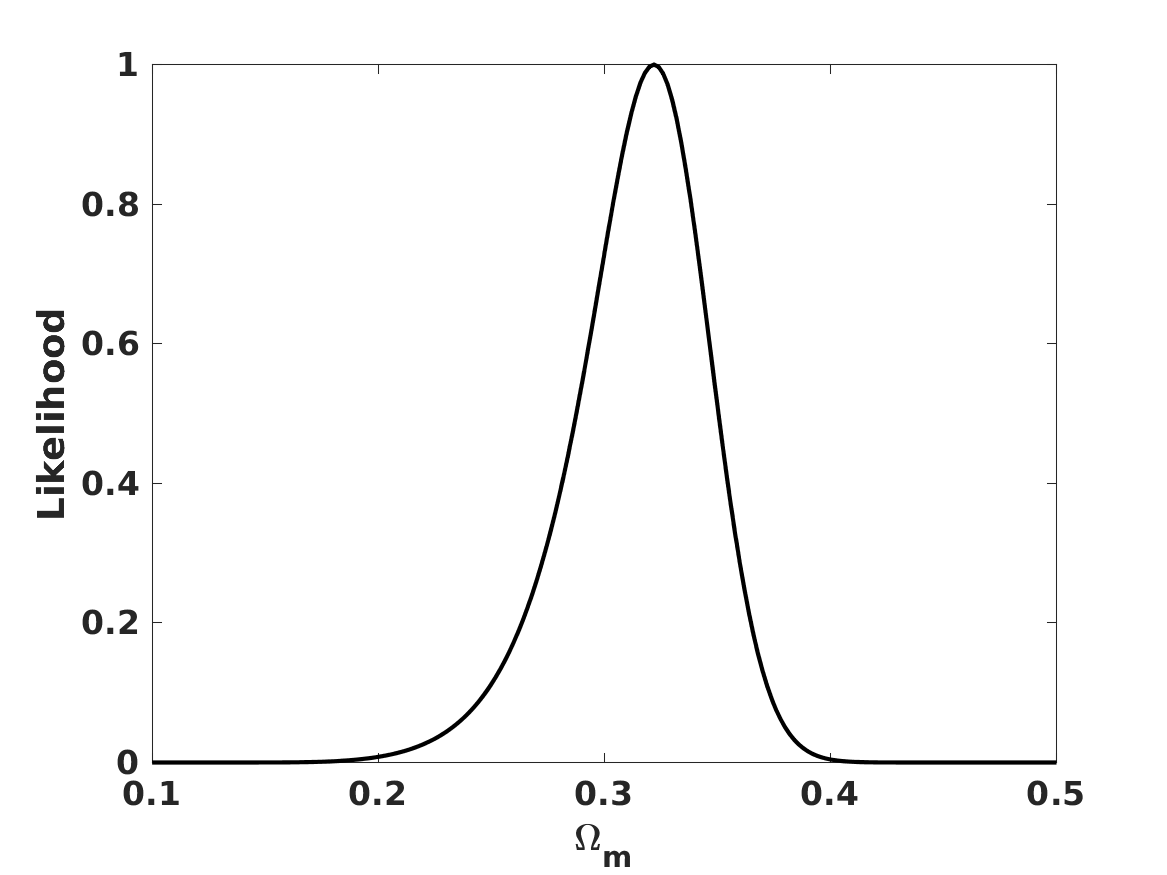}
\end{center}
\caption{\label{fig5}Posterior likelihood for the matter density in the $w=0$ Maeder model, for the combination of supernova and Hubble parameter data. The left panel depicts the result for $\Omega_k=0$, while in the right one $\Omega_k$ has been allowed to vary and marginalized.}
\end{figure*}

We start again with the $w=0$ case, and note that we can write $\Omega_\lambda=1-\Omega_m-\Omega_k$. The results of the likelihood analysis for this case are shown in Figure \ref{fig5}. If we further assume $\Omega_k=0$, the one-sigma posterior constraint in the matter density is
\begin{equation}
\Omega_m=0.26\pm0.02\,,\quad \chi^2_\nu=1.3\,,
\end{equation}
while if the curvature parameter is allowed to vary with the generous uniform prior $\Omega_k=[-0.2,0.2]$ and marginalized, one finds
\begin{equation}
\Omega_m=0.32\pm0.03\,,\quad \chi^2_\nu=1.2\,;
\end{equation}
it is clear from the structure of the Friedmann equation that the matter and curvature parameters are correlated. However, we note that the reduced chi-square at the best fit values is large in both cases, indicating that the model is not a good fit to the data, and further suggesting that a better solution can be found by widening the parameter space.

\begin{figure*}
\begin{center}
\includegraphics[width=3.2in,keepaspectratio]{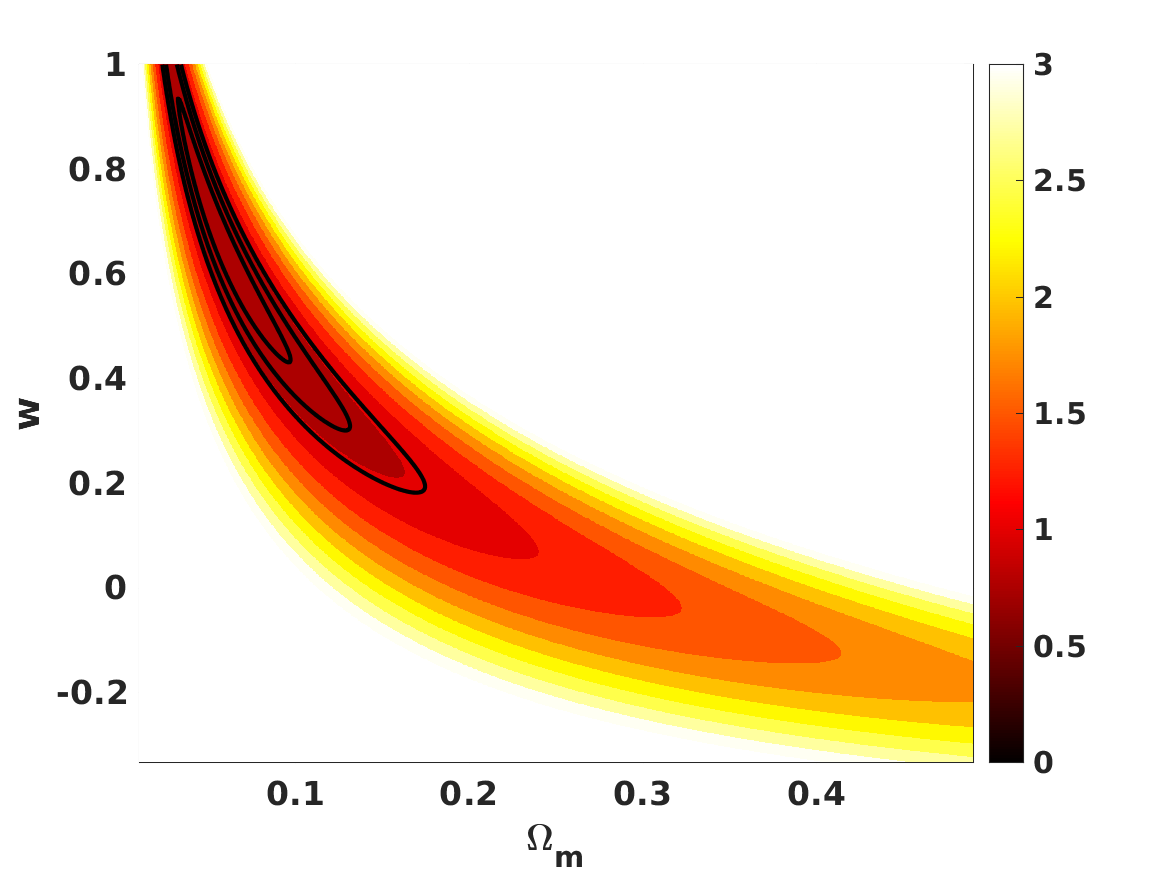}\\
\includegraphics[width=3.2in,keepaspectratio]{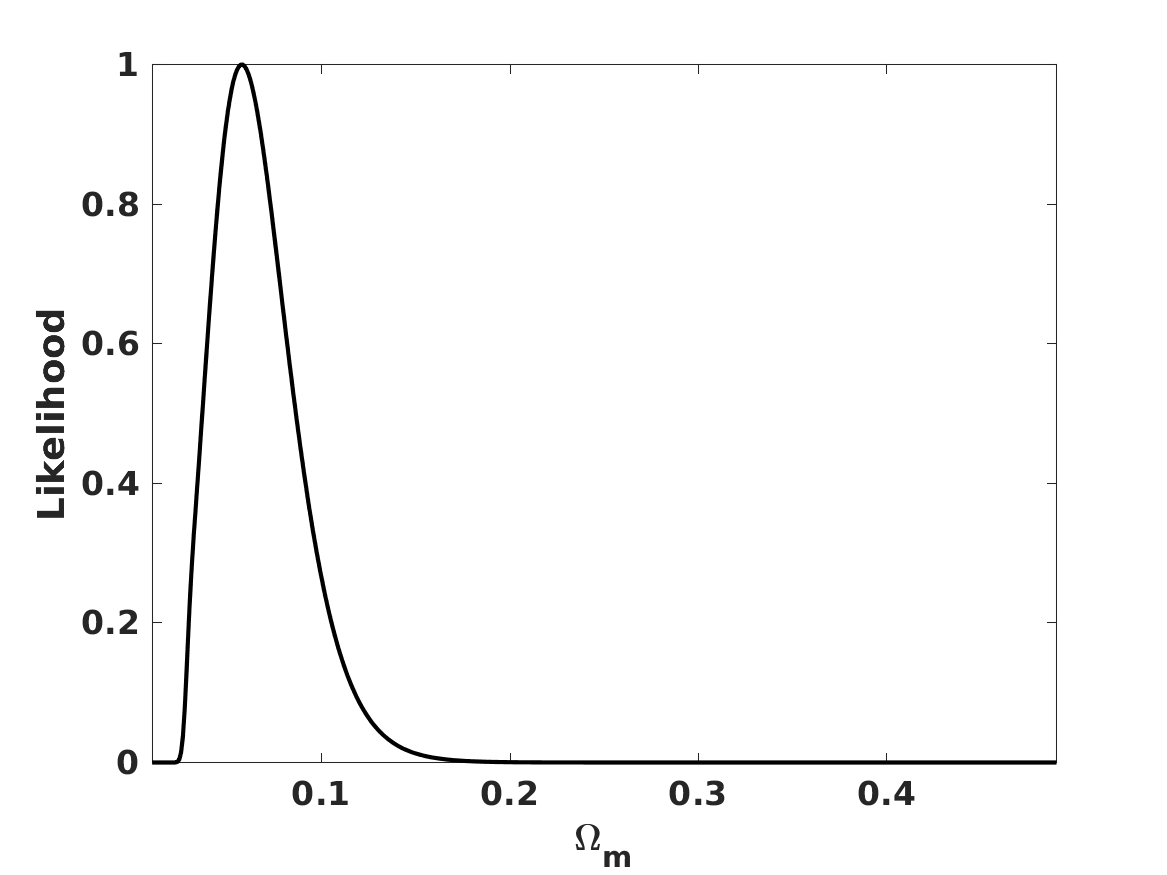}
\includegraphics[width=3.2in,keepaspectratio]{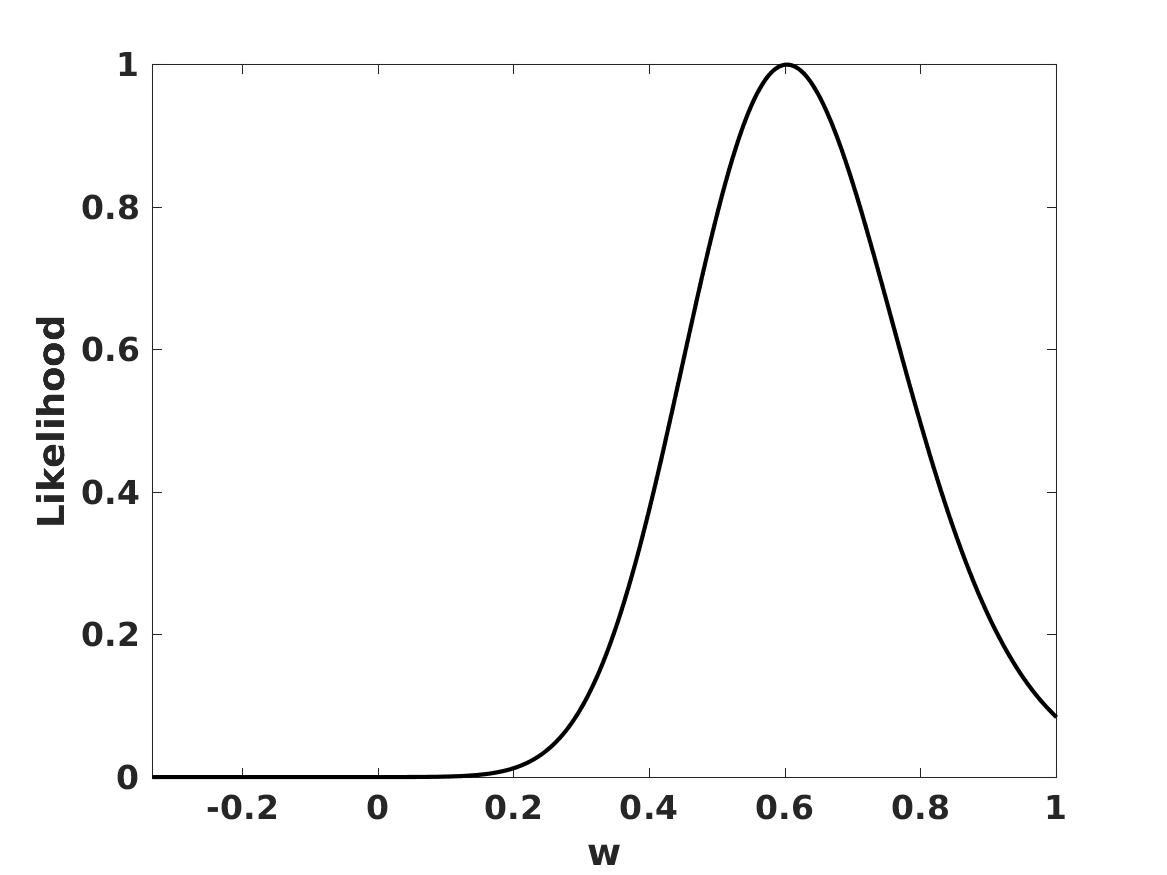}
\end{center}
\caption{\label{fig6}Two-dimensional and one-dimensional constraints (with the remaining parameters marginalized) for the Maeder model, with $\Omega_k=0$ and $w$ allowed to vary, for the combination of supernova and Hubble parameter data. In the two-dimensional case, the $\Delta\chi^2=2.3$, $\Delta\chi^2=6.17$ and $\Delta\chi^2=11.8$ confidence levels are shown in black lines, and the color map depicts the reduced chi-square at each point in the parameter space, with points with $\chi^2_\nu>3$ shown in white.}
\end{figure*}

This expectation can be readily confirmed by allowing $w$ as a further free parameter. The results of this analysis, for the $\Omega_k=0$ case, are shown in in Figure \ref{fig6}. In this case the one sigma constraints are
\bq
\Omega_m &=& 0.06\pm0.02\\ 
w &=&0.60^{+0.16}_{-0.15}\,,
\eq
while if $\Omega_k$ is allowed to vary (with the aforementioned prior) and marginalized one instead gets
\bq
\Omega_m &=& 0.06\pm0.03\\ 
w &=&0.59^{+0.17}_{-0.15}\,.
\eq
In both cases the reduced chi-square is now $\chi^2_\nu=0.8$, so the model is now slightly overfitting the data. Clearly there is a strong degeneracy between the matter density and the equation of state parameter, and the best fit values of both parameters are very far from the standard $\Lambda$CDM ones.

\section{Outlook}
\label{sect5}

We have compared three classes of models for the low-redshift acceleration of the universe against background low-redshift cosmological observations. Specifically, we used the traditional CPL phenomenological parametrization as a benchmark for the generalized coupling model of Feng and Carloni \cite{Feng} and the specific scale invariant model by Maeder \cite{Maeder}. Both of these can be interpreted as bimetric theories, but stem from very different underlying assumptions and, as we have seen, are subject to different observational restrictions.

In the generalized coupling model, the vacuum is by construction equivalent to General Relativity, but this equivalence is broken in the presence of matter. One practical consequence is that the model contains two types of vacuum energy: the usual cosmological constant plus a second contribution due to the matter fields, which effectively provides a new model parameter. On the other hand, the scale invariant model effectively has a time-dependent cosmological constant, but in the specific formulation of Maeder there is no parametric $\Lambda$CDM limit.

We start by noting that the in the CPL model the best-fit value of the matter density is $\Omega_m\sim0.26$ (which is slightly increased to $\Omega_m\sim0.27$ for the particular case of a constant equation of state parameter). For the generalized coupling model, it is clear that the usual cosmological constant is still necessary: setting it to zero would imply a matter density $\Omega_m>0.86$, in clear conflict with observations. Allowing for a cosmological constant, the model is therefore an extension of $\Lambda$CDM, with the additional parameter $Q$ describing the vacuum energy contribution due to the matter fields. In this case the best-fit value of the matter density increases by about one standard deviation (as compared to the CPL case), but there is no statistically significant evidence for a non-zero $Q$. Nevertheless, it is interesting to note that due to the quartic dependence of the Einstein equations on $Q$ there are two branches of the likelihood, with opposite  signs for $Q$. We have also checked that relaxing the standard assumption on the matter equation of state parameter, $w=0$, significantly broadens the posterior likelihoods but has a comparatively small impact on the best-fit parameters---in particular, $w=0$ itself is fully allowed.

The scale invariant model provides an interesting contrast. Assuming the standard equation of state parameter, $w=0$, the best-fit value is similar to the CPL one, though it increases if the curvature parameter is allowed (and marginalized). In any case, these fits have a relatively high reduced chi-square, indicating that the model does not provide a good fit to the data. Opening up the parameter space by allowing a non-zero (but still constant) equation of state parameter, one indeed finds a very different best-fit model, with $\Omega_m\sim0.6$ and $w=0.6$, with the two parameters being strongly anticorrelated.

From a purely phenomenological point of view, there are interesting analogies between these models and others that have been recently discussed in the literature. The generalized coupling model is similar to the steady-state torsion model of Kranas et al. \cite{Kranas}, recently constrained in \cite{Marques}, in that observations rule out their distinct physical mechanisms as the single source of the acceleration (in other words, a cosmological constant is still necessary), but still allow this mechanism to provide a small deviation from $\Lambda$CDM, with the corresponding dimensionless parameter being constrained at the percent level. On the other hand the scale invariant model is phenomenologically similar to the energy-momentum-powered models \cite{Roshan,Board,Akarsu}, recently constrained in \cite{Faria}, in the sense that the model in principle allows for a universe without a cosmological constant to be in reasonable statistical agreement with the data, although that would require values of the matter density and equation of state parameter that would disagree with other observations. Curiously, energy-momentum-powered models fit the data with a matter density that is slightly larger than the standard value and also prefer a slightly negative equation of state parameter, while in the scale invariant model the parameters shift in the opposite direction and do so more dramatically.

Finally, we note that each of the two signs of the matter vacuum energy parameter in the generalized coupling model corresponds to different physical interpretations and consequences, and it would be interesting to gain a better understanding of these. In particular, this could be translated into a physical prior in the analysis, possibly excluding one of the signs. As for the Maeder model, which is a specific example of a more general class of models of Canuto et al. \cite{Canuto1,Canuto2}, an open question is the extent to which the constrains apply  to the said general class. All in all, our analysis illustrates the point that alternative mechanisms to $\Lambda$CDM in explaining the low-redshift acceleration of the universe are very tightly constrained.

\section*{Acknowledgements}

This work was financed by FEDER---Fundo Europeu de Desenvolvimento Regional funds through the COMPETE 2020---Operational Programme for Competitiveness and Internationalisation (POCI), and by Portuguese funds through FCT - Funda\c c\~ao para a Ci\^encia e a Tecnologia in the framework of the project POCI-01-0145-FEDER-028987 and PTDC/FIS-AST/28987/2017. 

\bibliographystyle{model1-num-names}
\bibliography{models}
\end{document}